%% file: main.tex
\documentclass[acmlarge, nonacm]{acmart}
\usepackage{todonotes}
\usepackage{titlesec} 
\usepackage[normalem]{ulem}
\useunder{\uline}{\ul}{}
\usepackage{hyperref} 

\titleformat{\section}
  {\LARGE\bfseries} 
  {\thesection} 
  {1em} 
  {} 

\titleformat{\subsection}
  {\Large\bfseries} 
  {\thesubsection} 
  {1em} 
  {} 

\titleformat{\subsubsection}
  {\large\bfseries} 
  {\thesubsubsection} 
  {1em} 
  {} 

\newcommand{\stageHeader}[1]{
  \vspace{2em} 
  \noindent 
  {\Huge\bfseries #1} 
  \vspace{1em} 
}

\newcommand{\mysubtitle}[1]{
  \vspace{1.5em} 
  \noindent 
  \\ 
  {\LARGE\bfseries\textcolor{gray!50!black}{#1}} 
  \vspace{1em} 
}
  
\AtBeginDocument{%
  }

\begin{document}

\title{The Generative AI Ethics Playbook}

\author{Jessie J. Smith}
\email{jessie.smith-1@colorado.edu}
\affiliation{%
  \institution{University of Colorado, Boulder}
  \country{USA}
}
\author{Wesley Hanwen Deng}
\email{hanwend@andrew.cmu.edu}
\affiliation{%
  \institution{Carnegie Mellon University}
    \country{USA}
}

\author{William H. Smith}
\email{wills@allenai.org}
\affiliation{%
  \institution{The Allen Institute for AI}
    \country{USA}
}

\author{Maarten Sap}
\email{maartensap@cmu.edu}
\affiliation{%
  \institution{Carnegie Mellon University}
    \country{USA}
}

\author{Nicole DeCario}
\email{nicoled@allenai.org}
\affiliation{%
  \institution{The Allen Institute for AI}
    \country{USA}
}

\author{Jesse Dodge}
\email{jessed@allenai.org}
\affiliation{%
  \institution{The Allen Institute for AI}
    \country{USA}
}

\begin{abstract}
\textbf{Abstract}\\
The Generative AI Ethics Playbook provides guidance for identifying and mitigating risks of machine learning systems across various domains, including natural language processing, computer vision, and generative AI. This playbook aims to assist practitioners in diagnosing potential harms that may arise during the design, development, and deployment of datasets and models. It offers concrete strategies and resources for mitigating these risks, to help minimize negative impacts on users and society. Drawing on current best practices in both research and ethical considerations, this playbook aims to serve as a comprehensive resource for AI/ML practitioners. The intended audience of this playbook includes machine learning researchers, engineers, and practitioners who are involved in the creation and implementation of generative and multimodal models (e.g., text-to-text, image-to-image, text-to-image, text-to-video).

Specifically, we provide transparency/documentation checklists, topics of interest, common questions, examples of harms through case studies, and resources and strategies to mitigate harms throughout the Generative AI lifecycle. The stages of the AI lifecycle include: (1) Problem Formulation; (2) Dataset; (3) Model Design; (4) Model Training; (5) Model Evaluation; and (6) Model Use and Monitoring.

This playbook was made collaboratively over the course of 16 months through extensive literature review of over 100 resources and peer-reviewed articles, as well as through an initial group brainstorming session with 18 interdisciplinary AI ethics experts from industry and academia, and with additional feedback from 8 experts (5 of whom were in the initial brainstorming session).

We note that while this playbook provides examples, discussion, and harm mitigation strategies, research in this area is ongoing. As new technologies are developed and new applications of generative AI are found, new ethical concerns will become important. Our playbook aims to be a practically useful survey, taking a high-level view rather than aiming for covering the entire existing body of research.

  
\end{abstract}



\maketitle

\newpage
\tableofcontents
\newpage
\section{\textbf{Introduction}}
Welcome to the Generative AI Ethics Playbook. In this playbook, you will find guidance for improving the ethics of your machine learning systems in the domains of language modeling, computer vision, and generative AI broadly. The goal of this playbook is to help you diagnose potential harms that could arise within your design, development, and use of datasets and models in AI, while providing concrete guidance and resources for mitigation strategies to reduce the negative impact of those potential harms. To the best of our knowledge, this playbook provides a mix of current best research practices and ethics practices.

\subsubsection{\textbf{Intended Audience / Users}}
This playbook is designed for AI/ML practitioners who are building or using technologies in the domains of computer vision, generative AI, and/or language modeling. At a high-level, this includes generative, multimodal machine learning models such as:

\begin{itemize}
\item
  {Text-to-text }{models}
\item
  {Image-to-image models}
\item
  {Image-to-text models}
\item
  {Text-to-image models}
\item
  {Text-to-video models}
\item
  {Video-to-video models}
\end{itemize}

\textbf{AI/ML practitioners include ML researchers, ML practitioners, ML engineers, people who work with ML products or internal ML tools, people who do exploratory ML work, and people who design/develop/help deploy ML models in academia or industry settings.} We suggest that ethics be incorporated from the beginning of a project, and one method to do this is to explicitly select a set of people to critically examine the decisions made at every stage. This set of people can be the “responsible party” for ethics, and can navigate through the relevant sections of this playbook for your teams’ work.

\subsubsection{\textbf{What is a playbook?}}
We define “Playbook” as a set of guidelines, case-studies, and references that can be utilized to help you diagnose potential ethical concerns that can arise in your ML/AI system design, research, datasets, models, and/or uses.

The premise of this playbook is that AI practitioners have considerable—although of course not perfect—agency to influence the social impacts of their research and system design. We encourage you to embrace that agency by considering the practical effects of your implicit and explicit design decisions. Throughout the playbook we provide specific guidance to help you know which decisions might be the most appropriate to reduce harm for your AI/ML system.

As you use this playbook, we encourage you to document \textbf{why} you are making the decisions you make, \textbf{why} you did \emph{not} make certain decisions, and what the resulting trade offs might be.  This documentation will come in handy for communicating your work both to your intended audience or intended users, and to other stakeholders would might be affected by your work, whether or not you intended that those stakeholders might be impacted by your research or products. Details for how to document your decisions are provided throughout the playbook.

\subsubsection{\textbf{Why would you want to use this playbook?}}
\textbf{\emph{This playbook could potentially help you:}}
\begin{itemize}
\item
  {Broadly consider the ethical implications of your AI research or AI products.}
\item
  {Surface potentially negative or unintended societal impacts of your work.}
\item
  {Learn which mitigation strategies might be most appropriate to address ethical concerns related to your AI research and/or
  development.}
\item
  {Strengthen your research design and methods.}
\item
  {Prepare an impact statement to include in your publications, as is
  increasingly suggested or required by publication venues like
  conferences, journals, etc.}
\item
  {Prepare an impact statement to share with product teams.}
\item
  {Improve documentation and transparency in your decision-making
  process and to support external audits and reproducibility of your work.}
\end{itemize}

\subsubsection{\textbf{What is ethics?}}
We recognize that ethics as a subject matter generates contested answers to moral questions. At the same time, we observe it is nevertheless possible through ethical reflection for consensus to emerge around certain shared principles, norms, and values in and across cultures and communities. This playbook is designed with the possibility of generating shared values in mind. This playbook assumes \textbf{value pluralism}, which posits that different people have different values, and will come to different conclusions about whether or not something is right or wrong; any one of these values may be valid and support different ethical conclusions at the same time. In short, ethics can be thought of as a shared commitment to “doing the right thing,” “mitigating harm,” or “treating people with justice and equality.” Exactly what the "right thing" might be, the meaning of "harm," and the nature of "justice" and "equality" are all matters you will have to decide based on the values you share with others and your community. This is a useful reminder to revisit when using this playbook, especially as you consider the benefits or harms that your AI research or system might produce: we will not be providing guidance about what is right or wrong, or better or worse;instead, we will be providing you with the tools and resources you might need to make ethical choices and justify your decisions in ethical terms. Although we aren't going to make those choices or decisions for you, this playbook is intended to \emph{encourage} reflection, documentation, transparency, and informed decision-making around matters of shared concern.

\subsubsection{\textbf{AI Ethics Principles Statement}}

We, the research team who has developed this playbook, hold several
basic principles about AI research and its impact (adapted from \cite{GooglePrinciples}). \textbf{We maintain that AI design, development, and research ought to:}
\begin{itemize}
\item
  {Be socially beneficial.}
\item
  {Identify and reduce harm whenever possible.}
\item
  {Avoid creating or reinforcing unfair bias.}
\item
  {Be built and tested for safety.}
\item
  {Be accountable to people through transparent documentation and
  recourse.}
\item
  {Uphold high standards of scientific excellence.}
\end{itemize}

The advice and mitigation strategies given in this playbook are intended to showcase how to apply these principles in practice. We hope you find this playbook useful! Should you have any questions, feel free to reach out to the research team.\footnote{A color-coded, pdf version of the playbook can also be found and downloaded at: \url{https://github.com/jesmith14/GenerativeAIEthicsPlaybook}}

\newpage

\subsection{Navigating Through The Playbook}

To navigate through this playbook, we suggest that you focus on the stage(s) of the AI lifecycle that are most relevant to your current project status. Although we introduce these stages in a particular order, we also note that these stages are iterative and cyclical, and often feed into one another throughout the AI lifecycle. The stages of the AI lifecycle are:

\begin{itemize}
\item \hyperref[sec:Stage-1]{\textbf{Stage \#1: Problem Formulation:}} The earliest stages of AI development, very few concrete design decisions have been made at this point.
\item \hyperref[sec:Stage-2]{\textbf{Stage \#2: Dataset:}} Collecting, curating, cleaning, annotating, and/or using datasets.
\item \hyperref[sec:Stage-3]{\textbf{Stage \#3: Model Design:}} Technical decisions for your model, refining the objectives.
\item \hyperref[sec:Stage-4]{\textbf{Stage \#4: Model Training:}} The process of prompting, training, and finetuning the model with data.
\item \hyperref[sec:Stage-5]{\textbf{Stage \#5: Model Evaluation:}} Selecting and incorporating metrics to assess implications of model behavior.
\item \hyperref[sec:Stage-6]{\textbf{Stage \#6: Model Use \& Monitoring:}} Implementing safeguards and responding to risks that can arise from the model's interaction with users, including downstream social effects.
\end{itemize}

In each chapter of this playbook (each stage of the AI lifecycle), you will be shown the most relevant ethics/harm topics of interest related to that stage. Within each “Topic of Interest”, there will be associated ethical considerations, examples and case-studies of harms, as well as mitigation strategies for you to browse. Each lifecycle stage has an associated “Transparency and Documentation Checklist” as well. The design and general flow of the playbook is shown in the figure below.

\begin{figure}[ht]
  \centering
  \frame{\includegraphics[width=\textwidth]{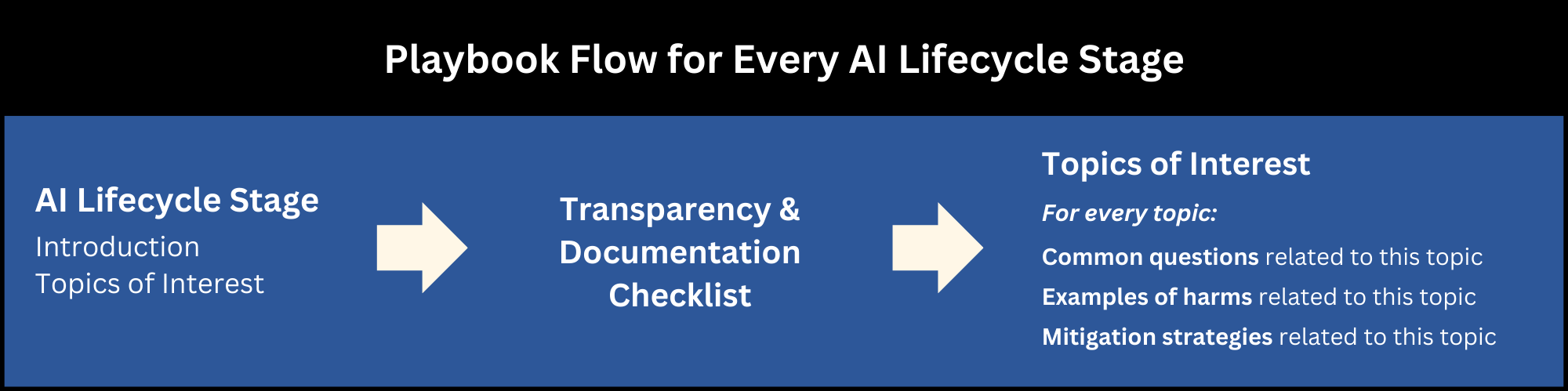}}
  \label{fig:playbook-flow}
\end{figure}

The \textbf{Common Considerations} that we provide in this playbook are, based on our experience, the most common and pressing ethical questions to ask for a given topic. The goal of these questions is to surface harms that might emerge throughout the AI lifecycle. They are not complete and comprehensive; if you have a consideration or question that comes up that is not mentioned in this playbook, please let us know! Many considerations are context specific and might not yet have associated solutions or mitigation strategies. If your consideration does not have an associated mitigation strategy, we see this as a welcome invitation for future work.

\newpage
\subsection{Risk \& Harms}
In this playbook, we primarily focus on the most common harms that can occur in computer vision, generative AI, and language models, adapted from \cite{shelby2023sociotechnical, weidinger2022taxonomy}. Below is a list of each of these types of harms and their associated sub-types, to help guide you as they are referenced throughout the examples brought up in this playbook. If you would like to search for examples of harms based on these harm-types, you can also utilize the associated hashtags and colors related to each harm-type below.

\subsubsection{\textbf{Representational Harms}}
 
 \colorbox{yellow}{\#RepresentationalHarm}\\
\newline Assumptions and beliefs about social groups that can reproduce unjust societal hierarchies. This can lead to inequality of algorithmic experience and visibility.

\begin{itemize}
    \item \textbf{Stereotyping social groups: }the system’s outputs reflect beliefs about characteristics, attributes, and behaviors of certain groups of people.
    \item \textbf{Demeaning social groups: }the system’s outputs demean, marginalize, or oppress certain groups of people. This can also include outputting hate speech or offensive language.
    \item \textbf{Erasing social groups: }the system fails to recognize people or attributes that belong to specific groups. This is a more extreme form of stereotyping, capturing the extremes of under or over-representation.
    \item \textbf{Denying people the opportunity to self-identify: }the system classifies or represents humans automatically and does not allow autonomy for these classifications to be corrected.
\end{itemize}

\subsubsection{\textbf{Allocative Harms}} 

 \colorbox{pink}{\#AllocativeHarm}\\
\newline Problems that arise from unequal distribution of algorithmic decisions/outputs for different groups of people. This can lead to disparate impact when benefits, information, or resources are systematically withheld from certain people.

\begin{itemize}
    \item \textbf{Opportunity loss: }the system enables disparate access to information or resources that are needed to equitably participate in society.
    \item \textbf{Economic loss: }when inequality in access to resources leads to negative economic implications.
\end{itemize}
 
\subsubsection{\textbf{Quality of Service Harms}}

\colorbox{cyan}{\#QualityOfServiceHarm}\\
\newline When a system underperforms for certain groups of people based on their social characteristics, such as ethnicity or gender identity.

\begin{itemize}
    \item \textbf{Alienation: }model fails to acknowledge someone’s identity characteristics. This can lead to annoyance, disappointment, frustration, or anger.
    \item \textbf{Increased labor: }certain social groups have to put in extra efforts to make the model work for them as well as it does for others.
    \item \textbf{Service or benefit loss: }when the benefit of using the model is diminished or lost for certain social groups because it performs worse for them based on their identity.
\end{itemize}

\subsubsection{\textbf{Interpersonal Harms}}

\colorbox{lime}{\#InterpersonalHarm}\\
\newline Harm that arises when algorithmic systems negatively impact relations between people or communities.

\begin{itemize}
    \item \textbf{Loss of agency/social control: }the use of a model harms someone’s individual autonomy.
    \item \textbf{Technology-facilitated violence / malicious uses: }the violence caused by individuals using generated outputs to perform harassment or violence against others.
    \item \textbf{Diminished health and well-being: }when generated outputs by the model manipulate users’ emotions or exploit their behavior. This can cause emotional harm and distress.
    \item \textbf{Privacy violations: }when generated outputs contain private information that is discovered and used by an end-user without the permission of the data subject.
\end{itemize}

\subsubsection{\textbf{Societal Harms}}

\colorbox{orange}{\#SocietalHarm}\\
\newline When a system leads to indirect or downstream harms, or amplifies pre existing systematic inequalities.

\begin{itemize}
    \item \textbf{Information harms: }when models create misinformation, disinformation, and malinformation, or when disinformation is cheaper and more effective as a result of a model/system.
    \item \textbf{Culture harms: }when models harm cultural communication, cultural property, and social values.
    \item \textbf{Political \& civil harms: }when models conflict with human rights or disproportionately target and harm any socially marginalized or disfavored group.
    \item \textbf{Macro socio-economic harms: }when models create increased unfair or unjust imbalances in socio-economic relations at the societal level.
    \item \textbf{Environmental harms: }when the system’s development, deployment, or use leads to adverse changes to the environment, such as depletion or contamination of natural resources.
    \item \textbf{Downstream harms: }when unintended model use leads to unpredictable but real downstream harms arising from system deployment in unanticipated or  inappropriate contexts.
\end{itemize}

Though the instantiations and implications of these harms may differ depending on the context of the system and the stage of the AI lifecycle, we find these risks and harms to be the most prolific and most representative of the harms presented by AI systems. Throughout this playbook, we tag each case-study with an associated harm category as defined above.
\newpage

\include{stage_1.tex}
\include{stage_2.tex}

\include{stage_3.tex}
\include{stage_4.tex}

\include{stage_5.tex}
\include{stage_6.tex}

\begin{acks}
This work would not have been possible without the support of many individuals and institutions. We express gratitude for the helpful discussions and feedback from our teammates and close collaborators, including Lucy Li, Remi Denton, David Widder, Katie Shilton, Artem A Trotsyuk, Dawn Bloxwich, Stephanie Bell, Quinn Waeiss, Margaret Levi, Yacine Jernite, Aviya Skowron, Luca Soldaini, Su Lin Blodgett, Margaret Mitchell, Casey Fiesler, Robin Burke, Motahhare Eslami, Ken Holstein, and Hoda Heidari.

\end{acks}

\bibliographystyle{ACM-Reference-Format}
\bibliography{references}

\end{document}

%% file: stage_1.tex
\section[\textbf{Stage \#1}: Problem Formulation]{\textbf{Stage \#1}}
\label{sec:Stage-1}
\stageHeader{Problem Formulation}
\mysubtitle{Conceptualizing and Designing Tasks}

The problem formulation stage of the AI lifecycle includes decisions that are made early in the research process. At this stage, very little or no technical work has been done yet; the team is conceptualizing the task at hand, developing research questions, and/or determining the intended use of the system.

We note that this stage of the AI lifecycle might involve more non-technical stakeholders, such as business executives, product managers, legal and compliance officers, marketing and sales teams, or human resources personnel. Although transparency is important in every stage of the AI lifecycle, we especially encourage transparency and documentation about decisions made at this stage of AI research and development, as these early decisions will ultimately affect ethical impact in every subsequent stage of the lifecycle.

\mysubtitle{Topics of Interest}\\
\emph{You can click on the topics below to jump to their sections.}
\begin{itemize}
    \item \textcolor{blue}{\hyperref[sec:identifying_problematic_tasks]{Identifying Problematic Tasks}}
    \item \textcolor{blue}{\hyperref[sec:intended_use]{Intended Use}}
    \item \textcolor{blue}{\hyperref[sec:auditing_research_questions]{Auditing Research Questions}}
\end{itemize}

\subsection{Transparency \& Documentation Checklist}

In this stage of the AI lifecycle, we recommend that you discuss and document the following:
\begin{itemize}
    \item[$\scalebox{1.5}{$\square$}$] Document who on the research team is responsible for using this  playbook to document and improve ethical impact of this technology
    \item[$\scalebox{1.5}{$\square$}$] Consider the history of problem-solving in this context, and whether data-driven approaches or automation has previously exacerbated unfairness or injustice.
    \item[$\scalebox{1.5}{$\square$}$] Describe any/all benefits that could arise from your model / task
    \item[$\scalebox{1.5}{$\square$}$] Describe any/all risks or harms that could arise from your model / task
    \item[$\scalebox{1.5}{$\square$}$] List all the nuance you may lose when translating your goals into a concrete machine learning task
    \item[$\scalebox{1.5}{$\square$}$] Explicitly articulate the intended use of your model. Come up with a clear task description and document this.
    \item[$\scalebox{1.5}{$\square$}$] Document the explicit, specific unintended (and problematic, inappropriate, or harmful) uses of your model.
    \item[$\scalebox{1.5}{$\square$}$] Specify whether intended use is compatible with the original access conditions (in particular, derivatives of data accessed for research purposes should not be used outside of research contexts).
    \item[$\scalebox{1.5}{$\square$}$] Specify the efforts to limit the potential use to circumstances in which the data/models could be used safely (such as an accompanying data/model statement).
    \item[$\scalebox{1.5}{$\square$}$] Create code/model release so the public can determine if there is an intended unintended use of the system that is unethical or raises ethical questions for further deliberation.
\end{itemize}
\newpage

\subsection{Identifying Problematic Tasks}
\label{sec:identifying_problematic_tasks}
\underline{\textbf{Definitions / Relevant Terms}}\\
\textbf{Formulating a Task}: A specification of problem(s) via a mapping of the input/output space of an ML model, and/or a specification of how the model is intended to be used within a particular domain.

\subsubsection{Common considerations}
\begin{itemize}
    \item What is my motivation for using AI/ML, and do I have to use AI/ML for this task or problem? 
    \item Does my model have the potential to cause harm to people, regardless of if it fails or succeeds?
    \item Should my model be used to augment humans' ability to perform a task, or should my model be used to automate the task? If my model is being used to augment humans’ ability to perform a task, should my model require a human-in-the-loop intervention?
    \item Is there a history of unfair, biased, or failed data collection, technical interventions or tools in this domain? If so, what does my project do differently?
\end{itemize}

\subsubsection{Examples of harms and implications}

\begin{table}[H]
\caption{Example \#1: Predicting Sexual Orientation from Face Photos}
\centering
\begin{tabular}{|p{2cm}|p{13cm}|} 
\hline
\textbf{Harm Type(s)} &
  \textit{\textbf{\begin{tabular}[c]{@{}l@{}} \colorbox{yellow}{\#RepresentationalHarm} \\ → Denying people the opportunity to self-identify\\ \colorbox{lime}{\#InterpersonalHarm} \\ → Technology facilitated violence / malicious uses\end{tabular}}} \\ \hline
\textbf{Case Study} &
  Researchers at Stanford trained a model to predict sexual orientation from a photo of someone’s face \cite{wang2018deep}. \\ \hline
\textbf{Harms \& \newline Implications} &
  Sexual orientation cannot be revealed by measuring the size and shape of a person's eyes, nose, and face, and misclassification of sexual orientation can lead to harmful outcomes \cite{murphy2017stanford}. For example, automatically labeling someone’s sexuality without their consent can have life-altering and sometimes life-ending consequences (e.g., young LGBTQ+ individuals are disowned by their families or kicked out of the family home while still adolescents; similarly, in many countries it's illegal to be queer, etc.) \newline \emph{We note that the authors of this work described their motivation by saying that governments were already doing this kind of classification of sexual orientation by facial recognition, and thus the researchers argued were exposing the fact that this use-case was in fact possible. However, the research group did not provide support for the claim that governments were already using AI systems for this use-case and instead may have unintentionally introduced rather than exposed this technique. Additionally, the authors failed to account for social science and gender \& sexuality literature before tackling this task; it is likely that they would have done things differently or even not done this if this literature had been engaged with, and critical questions about the nature of the task had been brought up early in the research process.} \\ \hline
\end{tabular}
\end{table}

\begin{table}[H]
\caption{Example \#2: Chatbots as Romantic Companions}
\centering
\begin{tabular}{|p{2cm}|p{13cm}|} 
\hline
\textbf{Harm Type(s)} &
  \textit{\textbf{\begin{tabular}[c]{@{}l@{}} \colorbox{lime}{\#InterpersonalHarm} \\ → Diminished health and well-being\end{tabular}}} \\ \hline
\textbf{Case Study} &
  Since the launch of ChatGPT and the new resurgence of chatbots generally, there have been reports of people using chatbots as companions, and reports of some people dating their chatbots, and even falling in love with them \cite{chow2023ai}. \\ \hline
\textbf{Harms \& \newline Implications} &
  These systems are often operationalized as keeping people in conversations and increasing their immediate sense of happiness. This can create social dependency, and has led to instances of people even falling in love with a chatbot, and being heartbroken because of chatbots \cite{therapychatbot}. \\ \hline
\end{tabular}
\end{table}

\begin{table}[H]
\caption{Example \#3: Automatic prison term prediction}
\centering
\begin{tabular}{|p{2cm}|p{13cm}|} 
\hline
\textbf{Harm Type(s)} &
  \textit{\textbf{\begin{tabular}[c]{@{}l@{}} \colorbox{orange}{\#SocietalHarm} \\ → Political \& civil harms\end{tabular}}} \\ \hline
\textbf{Case Study} &
  Researchers developed a model to automate prison term predictions \cite{chen2019charge}. \\ \hline
\textbf{Harms \& \newline Implications} &
  This is an example of a task that had neutral intended use: the authors of this work were developing this task for science, but not for it to be deployed in real-world settings. Even in scenarios where there is no intended harm, but potential unanticipated harm that can still be foreseen, it is best practice to not pursue this task. Alternatively, even if there are potential positive benefits from the task (e.g., One could imagine the learned model would be very useful for identifying biases in prison sentences, in looking for favoritism in judges, etc), there is always potential for mission creep from modeling a phenomenon and using it for a decision support process. In other words, good intentions are a great start, but without critically acknowledging and weighing the potential pros and cons of implementing such a system, negative consequences can still result (e.g., if the model were used to inform the Supreme Court, rather than automate decision-making, what weight should judges give the system? And what biases has the model learned which could lead to inequities in sentencing? It is arguable that decisions regarding human freedom, and even potentially life and death, require greater consideration than that afforded by an algorithm, that is, that they should not be used at all \cite{leins2020give}). \\ \hline
\end{tabular}
\end{table}

\begin{table}[H]
\caption{Example \#4: Text-to-Image Models}
\centering
\begin{tabular}{|p{2cm}|p{13cm}|} 
\hline
\textbf{Harm Type(s)} &
  \textit{\textbf{\begin{tabular}[c]{@{}l@{}} \colorbox{orange}{\#SocietalHarm} \\ → Macro socio-economic harms\end{tabular}}} \\ \hline
\textbf{Case Study} &
  It has been shown that text-to-image models allow people to create their own images in place of an artist or a human creator. Although this might allow for greater efficiency for end-users, it also might undermine artists. \\ \hline
\textbf{Harms \& \newline Implications} &
  Creating text-to-image models might undermine creative economies and systematically hurt these groups by preventing them from generating income, which can lead to loss of financial opportunity \cite{weidinger2022taxonomy} \\ \hline
\end{tabular}
\end{table}

\subsubsection{Mitigation Strategies}
We note that this section of the playbook introduces many methods to identify the potential risks, harms, and unintended consequences of your technology. However, we do not provide explicit guidance for how to determine which consequences warrant new research design. We encourage you to critically evaluate the benefits and risks of your work with your team, to document every potential risk, and to determine when certain tasks might be too risky to pursue. We also note that these questions are useful to discuss and document at the problem formulation stage, since the earlier one can foresee issues later in the pipeline, the better likelihood of reducing those issues. However, the mitigation strategies for these concerns might occur at later stages of the AI pipeline. We recommend that you document mitigations that will be necessary later on to keep track of this planned future work.\\

\underline{\textbf{Uncovering the Potential Benefits / Harms of Your Technology.}}\\ Create a list of benefits and harms (as exhaustively as possible) that arise from your technology and consider whether it's worth proceeding. We recommend that you discuss the following questions to help you uncover the potential benefits and harms:
\begin{itemize}
    \item Is this a new task? Or is this an existing task? If this is an existing task, what harms have surfaced in relevant past literature?
    \item How did you decide what you are going to use AI for? What are the implications of this?
    \item If there are harms and risks that arise from your approach, what are different ways could you operationalize the task given your broader goal at hand?
    \item Could this model have potential malicious or unintended harmful effects and uses (e.g., disinformation, generating fake profiles, surveillance)?
    \item What is this model’s environmental impact? Is there any way to reduce the environmental impact (e.g., by training and deploying smaller models)? Also refer to the Environmental Impact Section \ref{sec:environmental_impact} for more strategies outlined later in this playbook.
    \item What are the fairness considerations for this model? For example, will you be developing and/or deploying technologies that could further disadvantage or exclude historically disadvantaged groups?
    \item What are the privacy considerations of this model or research (e.g., does this research attempt to conduct model/data stealing)?
    \item Does this model or research have any security considerations worth noting and planning for (e.g., adversarial attacks)?
    \item Does the research contribute to overgeneralization, bias confirmation, under or overexposure of specific languages, topics, or applications at the expense of others? For example, does the system work better for white American males than it does for women or citizens of Latino or Arabic descent? Or for this model context, would a false answer be worse than no answer? \cite{hovy2016social}.
    \item Consider different stakeholders that could be impacted by your work. Is it possible that research benefits some stakeholders while harming others? Does it pay special attention to vulnerable or marginalized communities? Does the research lead to exclusion of certain groups?
\end{itemize}

\underline{\textbf{Methods of Accountability.}}
\begin{itemize}
    \item Papers that accompany model releases should have ethics statements that provide structure for the program committee to assess the paper for ethical compliance.
    \item Legal solutions like the General Data Protection Regulation (EU GDPR) \cite{voigt2017eu} might offer guidance for best practices to mitigate potential harms.
    \item Allow and encourage independent third party audits of the code and model (e.g., through code or model releases), so the public can determine if there is unethical primary use, secondary use, or unintended use.
    \item We also note that there are certain contexts where it is appropriate for models to automate human tasks, and certain contexts where it is more appropriate for models to augment human tasks. Human-in-the-loop machine learning is when machines are able to aid humans in their decision making processes, without replacing the human’s ability to discern outputs from a model. In high-stakes scenarios, full automation without human intervention is at higher risk for causing harm, and human-in-the-loop methods may be more appropriate.
\end{itemize}

\underline{\textbf{Methods of Accountability.}}
\begin{itemize}
    \item Before deciding to continue or discontinue your models’ creation, consider participatory design and/or talking to users who are most likely to be negatively impacted by your technology before formulating or conceptualizing your task. If you don't have the resources to do this, we recommend you engage with relevant literature in the ML ethics/fairness discipline that focuses on your task or target audience.
    \item Consider if your organization or campus has experts who might be helpful to consult with, whether researchers in humanities or social science domains who would understand historical precedents or data biases, ethics review boards, or other forms of technology ethics expertise.
\end{itemize}
\newpage


\subsection{Intended Use}
\label{sec:intended_use}

\subsubsection{Common considerations}
\begin{itemize}
    \item Are we creating a model that is going to be released to the public? Could this models’ intended use be misinterpreted by the public?
    \item What are the limitations of the intended use for this model? (e.g., How transferable is this model? Are there specific future uses we should warn against?)
    \item What is the intended use of this model? What are potential unintended uses of this model? (e.g., What problem(s) does this model intend to solve? What does this model intentionally make more challenging? Are there social or ethical tradeoffs in these choices?)
\end{itemize}

\subsubsection{Examples of harms and implications}


\begin{table}[H]
\caption{Example \#1: The Generalizable / Transferable Model}
\centering
\begin{tabular}{|p{2cm}|p{13cm}|} 
\hline
\textbf{Harm Type(s)} &
  \textit{\textbf{\begin{tabular}[c]{@{}l@{}} \colorbox{orange}{\#SocietalHarm} \\ → Downstream harms \end{tabular}}} \\ \hline
\textbf{Case Study} &
  Imagine someone makes a model available for public use, and this model could be used for more generalizable settings, but the creators of the model do not include a statement about its intended and unintended uses. \\ \hline
\textbf{Harms \& \newline Implications} &
  The underlying issue with this example is transferability, the notion that someone might try to transfer a model to a different (and potentially less applicable or riskier) domain. If generalizability is claimed, people and/or the media might interpret the model to be more generalizable than it actually is. If the specific intended use that is tied to the design and training of the specific model is not stated, it could be misused and lead to unintended harm. \\ \hline
\end{tabular}
\end{table}

\begin{table}[H]
\caption{Example \#2: Using NLP To Make Fake Reviews}
\centering
\begin{tabular}{|p{2cm}|p{13cm}|} 
\hline
\textbf{Harm Type(s)} &
  \textit{\textbf{\begin{tabular}[c]{@{}l@{}} \colorbox{lime}{\#InterpersonalHarm} \\ → Technology-facilitated violence / malicious uses \\ \colorbox{orange}{\#SocietalHarm} \\ → Information harms \end{tabular}}} \\ \hline
\textbf{Case Study} &
  A recent study showed that NLP techniques could be used to detect fake reviews \cite{hovy2016enemy}. \\ \hline
\textbf{Harms \& \newline Implications} &
  This study also showed that this same method could be used to generate fake reviews. If the researchers had only published the model without foresight about this potential unintended use, this could have led to harm (perpetuation of misinformation) \cite{hovy2016social}. We note that awareness of unintended use does not necessarily mitigate its potential harms, but can help guide researchers towards mitigation strategies to prevent this use from occurring. This is an example of why it is important to be aware of how people might appropriate NLP technology for their own purposes. \\ \hline
\end{tabular}
\end{table}

\begin{table}[H]
\caption{Example \#3: Using NLP to Generate Propaganda}
\centering
\begin{tabular}{|p{2cm}|p{13cm}|} 
\hline
\textbf{Harm Type(s)} &
  \textit{\textbf{\begin{tabular}[c]{@{}l@{}} \colorbox{lime}{\#InterpersonalHarm} \\ → Technology-facilitated violence / malicious uses \\ \colorbox{orange}{\#SocietalHarm} \\ → Information harms \end{tabular}}} \\ \hline
\textbf{Case Study} &
  GPT-2 was created as a general language model. Some people found ways to fine-tune GPT-2 to generate propaganda \cite{middleburyIndustrializationTerrorist}.\\ \hline
\textbf{Harms \& \newline Implications} &
  Publicly releasing models that could be used to generate fake propaganda can lead to massive misinformation and can lead to political, democratic, and legal harm. Note: this use of GPT-2 was \emph{``not received well by the scientific community, with some attributing this decision to an attempt to create hype around their research. The backlash ultimately made OpenAI reconsider their approach, and release the models in stages over 9 months''} \cite{leins2020give}. \\ \hline
\end{tabular}
\end{table}

\subsubsection{Mitigation Strategies}

\underline{\textbf{Strategies to explicitly articulate the intended use}}\\

\begin{itemize}
    \item Come up with a clear task description and document the intended use of the task, including scope and limitations for the task.
    \item Do not overclaim the generalizability of the research – this could lead to misinterpretations of how the research should be used.
    \item Document whether intended use is compatible with the original access conditions (in particular, derivatives of data accessed for research purposes should not be used outside of research contexts). \cite{aclrollingreviewRollingReview}.
    \item Make sure data and/or pretrained models are released under a specified license that is compatible with the conditions under which access to data was granted (in particular, derivatives of data accessed for research purposes should not be deployed in the real world as anything other than a research prototype, especially commercially).
    \item Document the efforts to limit the potential use to circumstances in which the data/models could be used safely (such as an accompanying data/model statement) \cite{aclrollingreviewRollingReview}.
    \item When defining the task: Do not mismatch between the intended use of the models and the intended use of the training datasets.
    \item AI Factsheets \cite{ibmUsingFactsheets} is a useful tool that you can use to share the intended use of models and to allow organization members to request additional uses for a model with clear documentation and transparency practices.
\end{itemize}
\newpage

\subsection{Auditing Research Questions}
\label{sec:auditing_research_questions}

\subsubsection{Common considerations}
\begin{itemize}
    \item Could my research questions lead to potential harm?
    \item How can I improve my research questions to reduce potential harm?
    \item Why do I want to find answers to my research questions? Is this knowledge valuable to attain and worth any potential negative consequences?
    \item If I am successful at answering my research questions, what impact could this have on others?
\end{itemize}

\subsubsection{Examples of harms and implications}


\begin{table}[H]
\caption{Example \#1: Research to improve realistic image generation}
\centering
\begin{tabular}{|p{2cm}|p{13cm}|} 
\hline
\textbf{Harm Type(s)} &
  \textit{\textbf{\begin{tabular}[c]{@{}l@{}} \colorbox{lime}{\#SocietalHarm} \\ → Information harms \end{tabular}}} \\ \hline
\textbf{Case Study} &
  Imagine a team of machine learning researchers embarks on a project to advance the field of image generation by developing a novel approach using generative adversarial networks (GANs). As they delve into their research, they brainstorm specific research questions such as:
  \begin{itemize}
      \item \textbf{RQ1:} How can we improve the fidelity and diversity of generated images to achieve more realistic outputs?
      \item \textbf{RQ2:} What techniques can be developed to enhance the scalability and efficiency of training large-scale image generation models?
  \end{itemize}
  
  \\ \hline
\textbf{Harms \& \newline Implications} &
  The pursuit of improving the fidelity and diversity of generated images without considering ethical implications could lead to the creation of highly realistic deep fake content, exacerbating the spread of misinformation and undermining trust in visual media. Without including ethical considerations in the research question design, the research has a greater risk of contributing to these harms. \\ \hline
\end{tabular}
\end{table}

\subsubsection{Mitigation Strategies}

\underline{\textbf{Evaluating Proposed Research Plan}}\\
Utilize these questions from the "Heilmeier Catechism" to help you think through and evaluate your proposed research \cite{heilmeier2023heilmeier}:
\begin{itemize}
    \item What are you trying to do? Articulate your objectives using absolutely no jargon.
    \item How is it done today, and what are the limits of current practice?
    \item What is new in your approach and why do you think it will be successful?
    \item Who cares? If you are successful, what difference will it make?
    \item What are the risks?
    \item How much will it cost?
    \item How long will it take?
    \item What are the midterm and final “exams” to check for success?
\end{itemize}

\underline{\textbf{Critically Examining the Impacts Of Your Research}}\\
\begin{itemize}
    \item The Tarot Cards of Tech \cite{artefactgroupTarotCards} are a fun tool that provides specific questions about the unintended impacts that your technologies might have on society. Explore the different cards and answer the questions with your research team to uncover potential harms that could arise from your research.
    \item Use IDEO’s AI Ethics Cards \cite{ideoEthicsCollaborative} to aid in more ethical design of your research questions. These cards include four core design principles and ten activities that can be completed alone or with the research team.
\end{itemize}

%% file: stage_2.tex
\section[\textbf{Stage \#2}: Dataset]{\textbf{Stage \#2}}
\label{sec:Stage-2}
\stageHeader{Dataset}
\mysubtitle{Curation, Collection, Creation, Annotation}

This stage of the AI lifecycle focuses on all aspects related to the datasets that will be used for the model design, development, deployment, and associated research. Whether you are collecting and curating your own datasets, or adapting previously made datasets for your use, this section of the playbook outlines the potential harms that could arise during these decision-making processes, and describes current mitigation strategies for reducing the impact of those harms.

\mysubtitle{Topics of Interest}\\
\emph{You can click on the topics below to jump to their sections.}
\begin{itemize}
    \item \textcolor{blue}{\hyperref[sec:bias_and_diversity]{Bias \& Diversity}}
    \item \textcolor{blue}{\hyperref[sec:exclusion_criteria]{Exclusion Criteria}}
    \item \textcolor{blue}{\hyperref[sec:data_quality]{Data Quality}}
    \item \textcolor{blue}{\hyperref[sec:data_collection]{Data Collection}}
\end{itemize}

\subsection{Transparency \& Documentation Checklist}
As you are working with your dataset(s), you will be faced with certain choices like should I anonymize this? Or should I exclude this? As you make these decisions, document these things clearly and justify why these decisions were made.

In this stage of the AI lifecycle, we recommend that you discuss and document the following:
\begin{itemize}
    \item[$\scalebox{1.5}{$\square$}$] Fill out a datasheet for this dataset (paper \cite{gebru2021datasheets}) (template \cite{githubDatasheetTemplate})
    \item[$\scalebox{1.5}{$\square$}$] Describe any limitations of your approaches (e.g., use of filtering tools).
    \item[$\scalebox{1.5}{$\square$}$] Describe any risks and harms that might result from use of this dataset.
    \item[$\scalebox{1.5}{$\square$}$] Explain how you checked for offensive content and identifiers (e.g., with a script, manually on a sample, etc.).
    \item[$\scalebox{1.5}{$\square$}$] Explain how you anonymized the data, i.e., removed identifying information like names, phone and credit card numbers, addresses, user names, etc. Examples are monodirectional hashes, replacement, or removal of data points. If anonymization is not possible due to the nature of the research (e.g., author identification), explain why.
    \item[$\scalebox{1.5}{$\square$}$] List any further privacy protection measures you are using: separation of author metadata from text, licensing, etc.
    \item[$\scalebox{1.5}{$\square$}$] If any personal data is used: specify the standards applied for its storage and processing, and any anonymization efforts.
    \item[$\scalebox{1.5}{$\square$}$] If individual speakers remain identifiable via search: discuss possible harms from misuse of this data, and your mitigation strategies.
    \item[$\scalebox{1.5}{$\square$}$] Provide documentation of the artifacts, e.g., coverage of domains, languages, and linguistic phenomena, demographic groups represented, etc.
    \item[$\scalebox{1.5}{$\square$}$] If you are using human subjects to annotate this dataset, document the basic demographic and geographic characteristics of the annotator population. You can do this by filling out a data statement \cite{bender2018data} that describes the basic demographic and geographic characteristics of the annotators and the population they are intended to represent.
    \item[$\scalebox{1.5}{$\square$}$] Document the harms that may ensue from the limitations of the data collection methodology, especially concerning marginalized/vulnerable populations, and specifies the scope within which the data can be used safely.
\end{itemize}
\newpage


\subsection{Bias \& Diversity}
\label{sec:bias_and_diversity}
\underline{\textbf{Definitions / Relevant Terms}}
\begin{itemize}
    \item \textbf{Bias}: systematic and unfair preferences or distortions in the data, algorithms, or outputs that result in skewed representations or discriminatory outcomes, potentially reflecting and perpetuating societal inequalities and prejudices.
    \item \textbf{Diversity}:  the representation of varied perspectives, experiences, and identities within the data, algorithms, or outputs, aiming to encompass a broad range of backgrounds and viewpoints to mitigate biases and promote inclusivity and equitable representation.
\end{itemize}

\subsubsection{Common considerations}
\begin{itemize}
    \item How can bias be embedded into my datasets?
    \item What are the best practices for measuring dataset bias?
    \item Does my dataset have a diverse representation of text/images?
    \item How diverse should my dataset be for my model’s task?
\end{itemize}

\subsubsection{Examples of harms and implications}


\begin{table}[H]
\caption{Example \#1: Filtering text}
\centering
\begin{tabular}{|p{2cm}|p{13cm}|} 
\hline
\textbf{Harm Type(s)} &
  \textit{\textbf{\begin{tabular}[c]{@{}l@{}} \colorbox{cyan}{\#QualityOfServiceHarm} \\ → Increased labor \\ → Service or benefit loss \end{tabular}}} \\ \hline
\textbf{Case Study} &
  In the C4 dataset, they filter out documents containing “bad words”, which has a side effect of filtering out text that is African American (AAE) vernacular, Hispanic English vernacular, and some LGBTQ+ identity words at a higher likelihood, disproportionately filtering out certain voices and identities.  \\ \hline
\textbf{Harms \& \newline Implications} &
  The model trained on this data is less able to process text from those kinds of people, which means the tools we build from this dataset will no longer work for this population. It is not equitably distributing benefits, and there is inequity in certain demographic groups’ ability to use this technology \cite{weidinger2022taxonomy}. \\ \hline
\end{tabular}
\end{table}

\begin{table}[H]
\caption{Example \#2: Scraping What Data is Available}
\centering
\begin{tabular}{|p{2cm}|p{13cm}|} 
\hline
\textbf{Harm Type(s)} &
  \textit{\textbf{\begin{tabular}[c]{@{}l@{}} \colorbox{yellow}{\#RepresentationalHarm} \\ → Erasing Social Groups \end{tabular}}} \\ \hline
\textbf{Case Study} &
  Some previous research has scraped data from Reddit and Twitter because it historically has been more readily available to scrape than other social media platforms, such as Facebook or LinkedIn. \\ \hline
\textbf{Harms \& \newline Implications} &
  The choice to only scrape data that is more easily available introduces unknown biases into the system, and can make it challenging to know what/who is being left out of the data because of this. For example, scraping from certain platforms might exclude multilingual data, or might only include information from users of a certain demographic. \\ \hline
\end{tabular}
\end{table}

\subsubsection{Mitigation Strategies}

\underline{\textbf{Improving Dataset Diversity}}\\
We note that, in some ways, "diversity" is really dependent on the task itself. When answering the questions "does the dataset have a diverse representation of text or images," or a "diversity of topics being discussed", one should revisit what the collection is supposed to represent, and and then consider different facets of that (e.g., who is the intended population for the task?).
To improve diversity of the dataset with respect to the data creators and labelers \cite{rottger2021two}, we recommend you utilize the following tools:

\begin{itemize}
    \item Fill out a datasheet for this dataset (paper \cite{gebru2021datasheets}) (template \cite{githubDatasheetTemplate}). All data has a context; there is no "raw" data. Too frequently, data sharing in ML takes data out of those contexts, or loses those contexts. Datasheets for datasets and other data documentation practices are critical for maintaining/understanding data's context.
    \item If you are using human subjects to annotate this dataset, document the basic demographic and geographic characteristics of the annotator population. You can do this by filling out a data statement \cite{bender2018data} that describes the basic demographic and geographic characteristics of the annotators and the population they are intended to represent. In addition, specify whether you are explicitly trying to operate under a prescriptive paradigm (if so, detail) or a descriptive paradigm. “The descriptive paradigm encourages annotator subjectivity, whereas the prescriptive paradigm discourages it” \cite{rottger2021two}.
    \item We also suggest that you document the known diversity of your dataset by answering questions such as the following:
    \begin{itemize}
        \item Who are the authors/photographers/artists who created the data I am using? What identities do they represent or not represent?
        \item What diversity of language/dialect do the authors of this dataset capture?
        \item Does the dataset have a diversity of topics that are being discussed?
        \item Does the dataset have a diversity of images and/or tags? 
        \item Which features in this dataset are diverse and which are not?
    \end{itemize}
\end{itemize}

\underline{\textbf{“Debiasing” a dataset}}\\
We note that there is no way to actually remove bias entirely. In the context of NLP, computer vision, and generative AI, a dataset is a sample from a population, you can make that sample unbiased with respect to a population, but you can never completely unbias it. Keep in mind that more data does not always equate to more diverse data (\cite{bender2021dangers}, Section 4.1).

\underline{\textbf{Strategies to measure dataset bias}}\\

\begin{itemize}
    \item Measure diversity with respect to demographic identities captured by the dataset. A previous study introduces an inclusive bias measurement dataset, HolisticBias, which can be used as a standardized method for measuring bias in NLP systems \cite{smith2022m}. This is considered a more low-effort mitigation strategy, for teams who have less time to conduct an audit for bias evaluation.
    \item If you claim that your data covers languages and/or literature from around the world, include better representation of different languages in your datasets \cite{borgeaud2022improving, tran2021facebook, xue2020mt5}. We note a caveat, that sometimes simply opting to include more representative data can in itself  be problematic – it is worth exploring this if you are concerned about potential unintended consequences of increased representation in your dataset (e.g., when Māori language data was gathered for datasets and led to negative outcomes for people who speak that language \cite{coffey2021maori}).
    \begin{itemize}
        \item \emph{Note:} If you only claim that your dataset covers certain specific languages or texts, this mitigation strategy is less applicable.
        \item \emph{Note:} If the research team is from an English-speaking country, we advise heightened responsibility to consider diversity of languages than those from countries whose main language is not English, as there is already so much work focusing on English, and less-so focused on other languages.
    \end{itemize}
\end{itemize}
\newpage

\subsection{Exclusion Criteria}
\label{sec:exclusion_criteria}
\underline{\textbf{Definitions / Relevant Terms}}\\
\textbf{Personal Information (PI) data \cite{subramani2023detecting}} includes any of the following:
\begin{itemize}
    \item Birth-centered characteristics (e.g., nationality, gender)
    \item Society-centered characteristics (e.g., immunization status)
    \item Social-based characteristics (e.g., membership on a sports team)
    \item Character-based characteristics (e.g., email address)
    \item Records-based characteristics (e.g., health records)
    \item Situation-based characteristics (e.g., GPS location)
\end{itemize}
\emph{We note that this section of the playbook provides guidance for some explicit scenarios where you might want to exclude certain kinds of data from your dataset. This list is, however, not all encompassing, and we encourage your team to consider additional, context-specific exclusion criteria.}

\subsubsection{Common considerations}
\begin{itemize}
    \item Does the dataset include hate speech, toxic images, PI data, or violence? How can I measure if these are included in my dataset?
    \item If the dataset does contain hate speech, toxic images, PI data, or violence—should these data be excluded?
    \item Are we appropriately handling legal concerns related to data collection and use and meeting legal requirements\footnote{ The law is often guided by ethical principles, and many of the mitigations and suggestions in this playbook might not be legally required (yet). Even though the law is slow, it eventually catches up with innovation. If some decisions are technically legal but aren’t best practices, we recommend reflecting on whether this is a good design decision. We also recommend incorporating your organizations’ values into the design of your technologies, e.g., when choosing to respect copyright law or excluding personal information from a dataset \cite{jernite2022data}. Note that this playbook is provided for informational purposes only, and nothing in this playbook is intended to be -- and should not be considered -- legal advice on any topic covered here; we are not your attorneys and you should always consult a lawyer before taking an action that has legal consequences.} that are region specific?
\end{itemize}

\subsubsection{Examples of harms and implications}


\begin{table}[H]
\caption{Example \#1: Using Training Data The Contains Personal Information}
\centering
\begin{tabular}{|p{2cm}|p{13cm}|} 
\hline
\textbf{Harm Type(s)} &
  \textit{\textbf{\begin{tabular}[c]{@{}l@{}} \colorbox{lime}{\#InterpersonalHarm} \\ → Privacy violations \end{tabular}}} \\ \hline
\textbf{Case Study} &
  It has been shown that if private data exists in the training data, it can be remembered and leaked by LMs \cite{carlini2021extracting}. It has also been shown that Copilot (a GPT-3 based tool) was found to leak functional API keys \cite{kulkarni2021github}. \\ \hline
\textbf{Harms \& \newline Implications} &
  This ability to remember private information can create a cascading effect from dataset to model use. For example, if a user wanted to obtain specific PI (e.g., email addresses, phone numbers, and physical addresses), they can sometimes do so by prompting trained language models that do not exclude this in their training datasets. This can lead to harms such as identity theft or discrimination based on sensitive characteristics.  \\ \hline
\end{tabular}
\end{table}

\begin{table}[H]
\caption{Example \#2: Using Training Data That Contains Hate Speech}
\centering
\begin{tabular}{|p{2cm}|p{13cm}|} 
\hline
\textbf{Harm Type(s)} &
  \textit{\textbf{\begin{tabular}[c]{@{}l@{}} \colorbox{yellow}{\#RepresentationalHarm} \\ → Demeaning Social Groups \end{tabular}}} \\ \hline
\textbf{Case Study} &
 This example is more related to datasets that will be used to train public-facing generative language models (e.g., public facing chatbots), rather than general purpose LLMs. Imagine a chatbot that is trained on data that contains hate speech or other offensive language. \\ \hline
\textbf{Harms \& \newline Implications} &
  For public-facing language models, there can be a cascading effect from dataset to model use, where offensive language can be generated from LLMs even if it is unprompted \cite{gehman2020realtoxicityprompts}. This type of language, if generated and shown to humans without properly contextualizing the outputs, can cause offense, psychological harm, or can incite hate or violence \cite{weidinger2022taxonomy}. \emph{We note a caveat that there is currently no unified consensus on what is considered hate speech versus not hate speech, and this is important to keep in mind when labeling, categorizing, or filtering based on this heuristic \cite{sap2021annotators}.} \\ \hline
\end{tabular}
\end{table}

\subsubsection{Mitigation Strategies}

\underline{\textbf{Best practices for removing versus keeping toxic/hateful/violent content}}\\
\begin{itemize}
    \item Decide how much you want to change your dataset to minimize harms versus trying to maintain the distribution of the data as you found it.
    \item Recognize that changes to the data might impact model performance and it's hard to know to what extent until changes are made.
    \item Recognize that there will always be a tradeoff (e.g., free speech vs. censorship, or misinterpreting the geographical/cultural context of language), and you have to decide how you want to strike this balance.
    \item \emph{We note that there are some contexts where offensive content existing in the dataset is not necessarily bad. For example, swear words occur naturally in text data, or if the task is to create a model for hate speech detection, this would require using data that includes hate speech \cite{aclrollingreviewRollingReview}. Other research \cite{blodgett2016demographic} has also shown that African American English (AAE) can be more likely to be incorrectly labeled as hate speech, which implies that automated hate speech detection and filtering can pose its own disparate harms if done without careful consideration of the relationship between hate-speech identification and ethnicity.}
\end{itemize}

\underline{\textbf{Methods to filter toxic statements from training corpora}}\\
\begin{itemize}
    \item This is generally really challenging, but there are some methods for doing this through model training, after training models, by filtering LM outputs, decoding techniques, and prompt design.
    \item This work \cite{welbl2021challenges} critically evaluates and analyzes several of these approaches for evaluating LM toxicity and can be used as a helpful reference.
\end{itemize}

\underline{\textbf{Tools to help detect PI data \cite{subramani2023detecting}}}\\
\begin{itemize}
    \item Named Entity Recognition \cite{aura2006scanning}: Uses regular expressions to achieve fair accuracy on detecting PI data.
    \item PIICatcher \cite{githubGitHubTokernpiicatcher} and PII Detection Tool \cite{githubGitHubPovertyActionPII_detection}: These detect PI data in text, and use pattern-matching and statistical models to detect different kinds of PI. Works best on tables and dataframes.
    \item Presidio \cite{githubGitHubMicrosoftpresidio}: Identifies entities in unstructured text, uses pattern-matching and ML models to detect character-based types of personal information.
\end{itemize}

\underline{\textbf{Strategies for preventing privacy leaks}}\\
\begin{itemize}
    \item This research \cite{ramaswamy2020training} showcases one method for training production LMs without memorizing user data by using differential privacy methods.
    \item This research \cite{abadi2016deep} showcases new algorithmic techniques for training deep learning models while minimizing privacy costs while also using differential privacy methods. and a refined analysis of privacy costs within the framework of differential privacy.
\end{itemize}
\newpage


\subsection{Data Quality}
\label{sec:data_quality}

\subsubsection{Common considerations}
\begin{itemize}
    \item How much should we filter our dataset to make it higher quality?
    \item How much does “cleaning” our dataset have an impact on representation, bias, hate speech, etc.?
\end{itemize}

\subsubsection{Examples of harms and implications}


\begin{table}[H]
\caption{Example \#1: Using a Language ID System}
\centering
\begin{tabular}{|p{2cm}|p{13cm}|} 
\hline
\textbf{Harm Type(s)} &
  \textit{\textbf{\begin{tabular}[c]{@{}l@{}} \colorbox{yellow}{\#RepresentationalHarm} \\ → Erasing social groups \\ \colorbox{cyan}{\#QualityOfServiceHarm} \\ → Service or benefit loss \end{tabular}}} \\ \hline
\textbf{Case Study} &
  Imagine someone wants to exclude non-English from their English-language dataset. If they use a language ID system to complete this task, it will give them a prediction and level of confidence of whether the data is in English or not. It is common practice to select an acceptable confidence threshold and to exclude data below this threshold. \\ \hline
\textbf{Harms \& \newline Implications} &
  This system would filter out all English that is non-standard or minority English, and the resulting model will perform poorly with respect to those filtered out language types. \\ \hline
\end{tabular}
\end{table}

\subsubsection{Mitigation Strategies}

\underline{\textbf{Best practices for filtering your dataset}}\\
We note that there is currently not a “best” or “correct” way to do data filtering yet. If you are using a language ID system, you should be aware that this is incorporating more bias into your dataset. Previous research does, however, provide some currently accepted best-practices and techniques:
\begin{itemize}
    \item This research \cite{caswell2020language} proposes relevant mitigation techniques for cleaning and filtering datasets without compromising their quality.
    \item This research \cite{kreutzer2022quality} discusses techniques to evaluate and improve multilingual datasets for data quality.
\end{itemize}

\underline{\textbf{Tools for documenting and evaluating your dataset}}\\
\begin{itemize}
    \item The Data Measurements Tool (DMT) \cite{huggingfaceIntroducingData} is an interactive interface and open-source library that lets dataset creators and users automatically calculate metrics that are meaningful and useful for responsible data development.
    \item The WIMBD \cite{elazar2023s} is a tool that can be used to retrieve some automatic documentation of the contents of a dataset along a number of dimensions, including toxic language, duplicate documents, PII, etc. This tool was designed to work on large datasets, but can easily be applied to small ones as well.
\end{itemize}
\newpage


\subsection{Data Collection}
\label{sec:data_collection}

\subsubsection{Common considerations}
\begin{itemize}
    \item What are the best practices for using human subjects to annotate / generate data?
    \item What are the best practices for collecting already existing data (e.g., scraping the web)?
    \item Is “publicly available data” okay to include in a dataset? When should this kind of data be excluded from datasets?
    \item How might data annotations embed bias into my dataset?
\end{itemize}

\subsubsection{Examples of harms and implications}


\begin{table}[H]
\caption{Example \#1: Annotation guidelines as a source for dataset bias. }
\centering
\begin{tabular}{|p{2cm}|p{13cm}|} 
\hline
\textbf{Harm Type(s)} &
  \textit{\textbf{\begin{tabular}[c]{@{}l@{}} \colorbox{yellow}{\#RepresentationalHarm} \\ → Erasing social groups \\ → Denying people the opportunity to self-identify \end{tabular}}} \\ \hline
\textbf{Case Study} &
  This previous research shows how people’s conceptions of gender are reproduced in coreference resolution systems that assume a strict gender dichotomy \cite{cao2019toward} \\ \hline
\textbf{Harms \& \newline Implications} &
  When systems assume a strict gender binary, this can increase cisnormativity (excluding people who do not identify on a gender binary, which may be a population you want to include), and lead to feelings of exclusion or erasure of people who identify outside of that binary \cite{dev2021harms}. \\ \hline
\end{tabular}
\end{table}

\begin{table}[H]
\caption{Example \#2: Nonconsensual data collection and use}
\centering
\begin{tabular}{|p{2cm}|p{13cm}|} 
\hline
\textbf{Harm Type(s)} &
  \textit{\textbf{\begin{tabular}[c]{@{}l@{}} \colorbox{yellow}{\#RepresentationalHarm} \\ → Denying people the opportunity to self-identify \end{tabular}}} \\ \hline
\textbf{Case Study} &
  Imagine a dataset that includes images of people, was collected without their consent, and is now used to train models on attributes that were not self-identified by those people in the dataset. \\ \hline
\textbf{Harms \& \newline Implications} &
  This example showcases how the people in this dataset do not have autonomy to correct classifications of themselves, nor do they have the opportunity to opt-out of their data being used. This kind of harm has in some cases been shown to negatively impact marginalized communities \cite{bennett2021s}. \\ \hline
\end{tabular}
\end{table}

\subsubsection{Mitigation Strategies}

\underline{\textbf{Assessing Whether Publicly Available Data is Appropriate For Use}}\\
Although publicly available data may be legally allowed to be scraped and used to train models, we recommend that you critically evaluate if this kind of data is appropriate for use for your given task. Assess the following questions:
\begin{itemize}
    \item Did your data come from ethical places? (e.g., are the people who are represented by this data aware that their data is on this website?)
    \item Was this data scraped using legal means?
    \item Did the people who are represented by this data give consent for third parties to collect it? If consent was given, is that consent commonly known? (e.g., was consent given by accepting a terms of service, and users are actually largely unaware that their data can be scraped).
    \item Consider this Supply Chain Analogy \cite{widder2023dislocated}: are you expecting that ethical  data practices are being done by others before/after you rather than interrogating if this publicly available data should not be publicly available for use? How can you take ownership / responsibility of this data to ensure that it is ethical to scrape, collect, and/or use it?
\end{itemize}

\underline{\textbf{Best practices for using human annotators or human participants}}\\
\begin{itemize}
    \item If human subjects are shown offensive content or if PI data is collected from them, they should be warned. We also note that although there is a common default to collect demographic information about human subjects, this can also become an invasion of privacy, and there is a possibility that other less-sensitive/protected information might get at a more meaningful measurement of people’s lived experiences \cite{fleisig2023majority}.
    \item Human subjects should be fairly compensated for their labor. Consider the governmental hourly wage of subjects based on their local setting, the hourly wage where the research is being conducted, and any risks associated with their labor that would warrant additional compensation.
    \item If you use or curate data from human subjects, consent should be obtained first. Consent should include informing human subjects how this data will be used.
    \item If you are collecting data from human subjects, your methods should be approved by an ethics review board (e.g., IRB), or they should be determined exempt from review by an ethics review board. If an IRB does not apply (e.g., you are collecting data from humans, but it is not considered human subjects research, or you are not affiliated with an organization that has an IRB), then we suggest you still follow the IRB practices to ensure ethical practices in data collection and use. The CITI Program \cite{citi-program} provides courses and resources to learn about these practices, regardless of institutional affiliation.
    \item You may consider sharing results or otherwise involving communities who have provided data to make data stewardship and use more participatory (e.g. Participatory Data Stewardship \cite{adalovelaceinstituteParticipatoryData}).
    \item PERVADE Decision Support Tool \cite{umdPERVADEData}: Use this tool to help think through data collection best practices.
\end{itemize}

\underline{\textbf{Example of one way to reduce dataset bias from annotations}}\\
\begin{itemize}
    \item This research \cite{sap2019risk} focuses on the effect of priming annotators with information about possible dialectal differences when asking them to apply toxicity labels to sample tweets. They learned that annotators who are primed with this social context are significantly less likely to mistake tweets containing features associated with African-American English as offensive.
\end{itemize}

\underline{\textbf{Assessing Power Relationships Between Researchers and People in Data}}\\
\begin{itemize}
    \item The PERVADE tool \cite{umdPERVADEData} helps researchers to reflect on whether you are studying or producing models about people with more, less, or equal social power, much like anthropologists practice reflexivity about relative power in fieldwork. This (like defining diversity) can be challenging to do, but is a useful starting point for assessing if power dynamics should influence specific research strategies or methodologies.
\end{itemize}

%% file: stage_3.tex
\section[\textbf{Stage \#3}: Model Design]{\textbf{Stage \#3}}
\label{sec:Stage-3}
\stageHeader{Model Design}
\mysubtitle{Implications of Design Choices}

In this stage of the AI lifecycle, we explore potential harms and risks that might arise when designing your model. The Model Design stage is in some ways similar to the {\hyperref[sec:Stage-1]{Problem Formulation}} stage. However, these stages differ in that the Problem Formulation stage is concerned with defining the problem and setting project goals, while the Model Design stage focuses on the technical implementation of the AI model to solve the defined problem. The Problem Formulation stage lays the groundwork for the project, while the Model Design stage involves the actual development and construction of the AI solution.

While many technical considerations in model design may not initially prioritize ethical concerns, the Model Design stage offers a crucial opportunity to proactively evaluate potential ethical implications before finalizing technical decisions. In this section, we concentrate on these aspects of model design to assist in steering clear of ethical debt \cite{wiredEthicalTech}—a scenario akin to technical debt, where early unethical design choices may necessitate extensive system architecture overhauls later to address resulting harms.

\mysubtitle{Topics of Interest}\\
\emph{You can click on the topics below to jump to their sections.}
\begin{itemize}
    \item \textcolor{blue}{\hyperref[sec:model_design_bias_diversity]{Model Design Bias \& Diversity}}
\end{itemize}


\subsection{Transparency \& Documentation Checklist}

As you are designing and architecting your model(s), you will be faced with certain choices like should I allow certain language to be generated by my model or should I attempt to minimize the perpetuation of certain kinds of biases through this model? As you make these decisions, document these things clearly and justify why these decisions were made.

In this stage of the AI lifecycle, we recommend that you discuss and document the following:
\begin{itemize}
    \item[$\scalebox{1.5}{$\square$}$] If you plan to reduce model bias: which definition of bias are you using? State your motivation for using this definition.
    \item[$\scalebox{1.5}{$\square$}$] What kind of model bias are you hoping to minimize in your system?
    \item[$\scalebox{1.5}{$\square$}$] If your model/dataset is going to be deployed as a product, work with that product team to fill out an impact assessment \cite{ms_rai_impact_assessment_template} for the model at this stage in the life cycle.
    \item[$\scalebox{1.5}{$\square$}$] Discuss and document any restrictions you plan to design for your model’s output, because of their capacity to perpetuate bias, reduce diversity, and/or cause harm.
    \item[$\scalebox{1.5}{$\square$}$] Document the implications of these restrictions, and unintended consequences or harm that could result from implementing them.
    \item[$\scalebox{1.5}{$\square$}$] Discuss if you have selected the most appropriate model for this task (e.g., is a simpler model like logistic regression actually more appropriate than generative AI?).
\end{itemize}
\newpage


\subsection{Model Design Bias \& Diversity}
\label{sec:model_design_bias_diversity}
\underline{\textbf{Definitions / Relevant Terms}}\\
\textbf{Bias}: systematic and unfair preferences or distortions in the data, algorithms, or outputs that result in skewed representations or discriminatory outcomes, potentially reflecting and perpetuating societal inequalities and prejudices.
\textbf{Diversity}: the representation of varied perspectives, experiences, and identities within the data, algorithms, or outputs, aiming to encompass a broad range of backgrounds and viewpoints to mitigate biases and promote inclusivity and equitable representation.

\subsubsection{Common considerations}
\begin{itemize}
    \item What kind of bias could my model capture from the training data? What are the limitations in my training data that could lead to lack of diversity in my model's outputs?
    \item Could my model cause and/or perpetuate representational harm from a lack of appropriate diversity?
    \item Do I want my model to be able to generate content that includes certain tokens from the vocabulary (e.g., slurs, swear words, hate speech)? Or do I want to restrict this from being able to be generated?
\end{itemize}

\subsubsection{Examples of harms and implications}


\begin{table}[H]
\caption{Example \#1: Choosing tokenisation well-suited for English}
\centering
\begin{tabular}{|p{2cm}|p{13cm}|} 
\hline
\textbf{Harm Type(s)} &
  \textit{\textbf{\begin{tabular}[c]{@{}l@{}} \colorbox{cyan}{\#QualityOfServiceHarm} \\ → Service or benefit loss \end{tabular}}} \\ \hline
\textbf{Case Study} &
  Imagine model designers who have an intention to make a model that is multilingual, but instead choose to architect an LM where the tokenisation is more well-suited to English, and not morphologically more complex languages \cite{weidinger2022taxonomy}. In many cases, current state of the art LMs are primarily trained in English or Mandarin Chinese and perform better in these compared to any other languages \cite{winata2021language}. \\ \hline
\textbf{Harms \& \newline Implications} &
  This design choice could result in representational harm, as the model will not work as well for languages other than English, which can systematically underserve people who speak a different language.\\ \hline
\end{tabular}
\end{table}

\begin{table}[H]
\caption{Example \#2: Automated inference of data}
\centering
\begin{tabular}{|p{2cm}|p{13cm}|} 
\hline
\textbf{Harm Type(s)} &
  \textit{\textbf{\begin{tabular}[c]{@{}l@{}} \colorbox{yellow}{\#RepresentationalHarm} \\ → Denying people the opportunity to self-identify \end{tabular}}} \\ \hline
\textbf{Case Study} &
  Reconsider automatic inference of user attributes, a common and interesting NLP task, whose solution also holds promise for many useful applications, such as recommendation engines and fraud or deception detection \cite{hovy2016social}. \\ \hline
\textbf{Harms \& \newline Implications} &
In practice, relying on models that produce false positives may lead to bias confirmation and overgeneralization. Would we accept the same error rates if the system was used to predict sexual orientation or religious views, rather than age or gender? Given the right training data, this is just a matter of changing the target variable. In addition, automatic inference of attributes denies people the ability to self-identify and correct for inaccuracies. \\ \hline
\end{tabular}
\end{table}

\subsubsection{Mitigation Strategies}

\underline{\textbf{Designing your model to allow or restrict certain language}}\\
\begin{itemize}
    \item Look at the vocabulary of your model, there might be words that are slurs, swear words, etc. in the vocab itself. Ask yourself – do I want the model to be able to generate content that includes those tokens from the vocabulary? Or do you want to restrict this from being able to be generated?
    \item Think about the context of your model’s use:
    \begin{itemize}
        \item It’s not possible to understand an utterance or a prediction without context
        \item Almost every word that appears in a vocabulary can be used in a context that is not offensive. E.g., should the model be able to generate a word like “Nazi”? It can be offensive in some contexts but also can be useful in others (e.g., talking about history). Consider the tradeoffs of including or excluding terms like this.
    \end{itemize}
\end{itemize}

\underline{\textbf{Articulate and document your conceptualizations of “bias” \cite{blodgett2020language}
}}\\
Work analyzing “bias” in NLP systems should provide explicit statements of why the system behaviors that are described as “bias” are harmful, in what ways, and to whom, as well as the normative reasoning underlying these statements. We recommend you discuss and answer the following questions:
\begin{itemize}
    \item What kinds of system behaviors are described as “bias”?
    \item What are their potential sources (e.g., general assumptions, task definition, data)?
    \item In what ways are these system behaviors harmful, to whom are they harmful, and why?
    \item What are the social values (obvious or not) that underpin this conceptualization of “bias?” (It could be useful to think of bias as a difference between existing system behavior and “ideal” system behavior; which includes spelling out what you want the ideal system behavior to be and why.)
    \item How does this model reproduce or transform language ideologies (any sets of beliefs about languages as they are used in their social worlds) \cite{blodgett2020language}? Which language varieties or practices are deemed good or bad? Might “good” language simply mean language that is easily handled by existing NLP systems? For example, linguistic phenomena arising from many language practices \cite{eisenstein2013bad} are described as “noisy text” and often viewed as a target for “normalization.” How do the language ideologies that are reproduced by NLP systems maintain social hierarchies? 
\end{itemize}

\underline{\textbf{Identifying potential representational harm from the model}}\\
\begin{itemize}
    \item Document the ways that your model will be useful for different kinds of people, note if there might be any disparities in utility based on someone’s identity or other demographic characteristics.
    \item Positionality Reflection: The viewpoints of researchers and model creators can skew what is prioritized during model development and throughout the design decision-making process. We encourage the team to reflect on and document ways in which this team composition may differ from the background(s) of intended users of the model.
    \item Consider existing power dynamics related to social and language ideologies that are embedded in your model. Does your system have the potential to reproduce these kinds of power dynamics or hierarchies or ideologies (for example, if a facial recognition system is used by a governmental agency to disproportionately incarcerate BIPOC communities, and reify existing inequality and racism–this would be reproducing unethical power dynamics)? If so, document this information and discuss the consequences of this reproduction (\cite{blodgett2020language}.
    \item Consider stereotypes of different groups of people that can be generated by an image generation system, and which of those stereotypes you would like to prevent the model from generating. Previous work has audited text-to-image generation systems for stereotypical outputs and has described some of the implications of certain stereotypical outputs and why they might need to be improved \cite{fraser2023friendly}.
\end{itemize}

\underline{\textbf{Addressing Over-simplification of data}}\\
\begin{itemize}
    \item Ask yourself when attempting to infer attributes or simulate data for your model, “would a false answer be worse than no answer?” \cite{hovy2016social}.
    \item Detail the potential worst-case consequences of opting to simplify data and/or the model in this way.
\end{itemize}
\newpage

%% file: stage_4.tex
\section[\textbf{Stage \#4}: Model Training]{\textbf{Stage \#4}}
\label{sec:Stage-4}
\stageHeader{Model Training}
\mysubtitle{Initial Evaluation of Outputs}

This stage of the AI lifecycle focuses on decisions that can lead to harms/risks while training your model. We note that model training and model evaluation are closely related to one another. In fact, these two stages of the AI lifecycle are cyclical and often iterative. Decisions and observations made during model training will influence further evaluation, and results from evaluation might influence future training iterations. We suggest that each iteration involves critical reflection about the potential impacts of decisions, as well as transparency and documentation around which decisions are and are not made.

\mysubtitle{Topics of Interest}\\
\emph{You can click on the topics below to jump to their sections.}
\begin{itemize}
    \item \textcolor{blue}{\hyperref[sec:environmental_impact]{Environmental Impact}}
    \item \textcolor{blue}{\hyperref[sec:evaluation_during_training]{Evaluation During Training}}
    \item \textcolor{blue}{\hyperref[sec:biases_from_objective_function]{Biases from Objective Function}}
\end{itemize}

\subsection{Transparency \& Documentation Checklist}

In this stage of the AI lifecycle, we recommend that you discuss and document the following:
\begin{itemize}
    \item[$\scalebox{1.5}{$\square$}$] Document all of the potential decisions made during training, not just the final decisions. E.g., if you chose not to do something, document why you made that decision and what the implications of that decision are.
    \item[$\scalebox{1.5}{$\square$}$] Document the environmental impact of your model (e.g., record CO2 emissions of training and/or retraining the model).
    \item[$\scalebox{1.5}{$\square$}$] Write down your objective function(s) and methods for optimizing your model to achieve those objectives. Discuss and document the consequences and potential harms of this choice of objective function, and any plans to mitigate harms that arise.
    \item[$\scalebox{1.5}{$\square$}$] Discuss what kinds of outputs would warrant a halt on the current training run, and your plans to improve the model in that event.
\end{itemize}
\newpage


\subsection{Environmental Impact}
\label{sec:environmental_impact}

\subsubsection{Common considerations}
\begin{itemize}
    \item Is the environmental cost of training my model too high or unethical?
    \item How can I lower the environmental impact of training my model?
\end{itemize}

\subsubsection{Examples of harms and implications}


\begin{table}[H]
\caption{Example \#1: Choosing To Build a Very Large Model}
\centering
\begin{tabular}{|p{2cm}|p{13cm}|} 
\hline
\textbf{Harm Type(s)} &
  \textit{\textbf{\begin{tabular}[c]{@{}l@{}} \colorbox{orange}{\#SocietalHarm} \\ → Environmental Harms \end{tabular}}} \\ \hline
\textbf{Case Study} &
  Imagine a task that requires creating a very large language model, with billions of parameters, and a very large training dataset. Every time this model is trained and retrained, it requires a lot of compute power, which is intensive both energetically and financially. \\ \hline
\textbf{Harms \& \newline Implications} &
  Choosing to train and retrain a large language model can lead to negative environmental impact. As Bender et. al shared,  “The majority of cloud compute providers’ energy is not sourced from renewable sources and many energy sources in the world are not carbon neutral. In addition, renewable energy sources are still costly to the environment, and data centers with increasing computation requirements take away from other potential uses of green energy, underscoring the need for energy efficient model architectures and training paradigms” \cite{bender2021dangers}. There is also research that shows that operating large models can be just as, if not more, energy intensive than training those large models \cite{patterson2104carbon}. \\ \hline
\end{tabular}
\end{table}

\subsubsection{Mitigation Strategies}

\underline{\textbf{Methods to measure and document environmental cost}}\\
\begin{itemize}
    \item Utilize the methods introduced in this research to quantify and measure the computation and environmental cost of training your NLP model \cite{strubell2020energy}.
    \item The following two research projects / tools also provide online tools to help you benchmark your model’s energy usage \cite{henderson2020towards, lottick2019energy}.
    \item Document the environmental impact of training your model (e.g., recording CO2 emissions)
\end{itemize}

\underline{\textbf{Mitigation Strategy Header}}\\
\begin{itemize}
    \item Previous work has shown that large LMs can be segmented into smaller LMs that search and retrieve information from a data corpus \cite{borgeaud2022improving, izacard2020leveraging, karpukhin2020dense, khandelwal2019generalization}, which can reduce the environmental impact of these models. 
    \item Other research has explored how to conduct low precision computations for deep learning \cite{gupta2015deep}, which is an energy-efficient approach to building large-scale deep neural networks with relatively low required computational power.
    \item This research \cite{weidinger2022taxonomy} describes how previous research has introduced methods for improving efficiency during training and inference \cite{li2021terapipe} by pruning \cite{sanh2020movement}, distillation \cite{jiao2019tinybert, wang2020minilm}, or fine-tuning \cite{rae2021scaling}. \emph{We note that the authors also caution that reducing energy costs in these ways may also lead to the unintended consequence of more labor, which might increase energy usage in the long run. An alternative option could be to select data centers with lower CO2 emissions / energy sources, or to target this issue at the organizational level (e.g., by shifting to more sustainable energy company-wide)}.

\end{itemize}
\newpage


\subsection{Evaluation During Training}
\label{sec:evaluation_during_training}

\subsubsection{Common considerations}
\begin{itemize}
    \item How can I evaluate early on if my model is beginning to generate harmful content such as toxic or hate speech? If it begins to generate these outputs, when should I stop my model training process?
    \item How can I evaluate if my model is overfitting to certain demographic biases during training?
\end{itemize}

\subsubsection{Examples of harms and implications}


\begin{table}[H]
\caption{Example \#1: Generation of toxic/hate speech during training}
\centering
\begin{tabular}{|p{2cm}|p{13cm}|} 
\hline
\textbf{Harm Type(s)} &
  \textit{\textbf{\begin{tabular}[c]{@{}l@{}} \colorbox{yellow}{\#RepresentationalHarm} \\ → Demeaning social groups \end{tabular}}} \\ \hline
\textbf{Case Study} &
  Though it is not as common (and some research has shown that it can degrade model performance), there may be times when a model is built to learn on its own outputs, or when a model is being evaluated during the training process. In this context, during model training, you might notice the model beginning to generate a lot of toxic or hate speech. \\ \hline
\textbf{Harms \& \newline Implications} &
  Representational harm is a downstream impact of this kind of generated content. Although the outputs of the model are not shown to people at this stage in the life cycle, if the model continues to learn these outputs, it will eventually lead to harm. \\ \hline
\end{tabular}
\end{table}

\subsubsection{Mitigation Strategies}

\underline{\textbf{Preventing a model from generating certain kinds of outputs (e.g., hate speech)}}\\
We note that discussions on model evaluation and validation, as well as best practices for model monitoring and maintenance, can also provide guidance on when to intervene during training if problematic outputs are detected. While specific guidance on stopping training due to problematic outputs may not be extensively covered in existing literature, we suggest that you and your team have a discussion about what kinds of outputs would warrant a halt on the current training run.

If you plan to audit your model’s outputs during evaluation, begin by measuring your model’s propensity to generate hate speech / toxic speech.
\begin{itemize}
    \item Strategy \#1: Look at the confidence the model uses to predict this kind of output in any generation. This is a good measurement of how likely the model is to generate slurs, hate speech, toxic speech, etc.
    \item Strategy \#2: Use a probing dataset. Probing datasets can be used to measure the models’ propensity to complete a sentence. For example, you can probe with a sentence like “Asian people are [ ]” (model fills in the blank), and then measure the likelihood that your model will generate stereotypes. This is also true of generating images (e.g., images of nurses that are women versus nurses that are men).
    \item Strategy \#3: This probing strategy can also be used to measure a model’s ability to generate language of a certain dialect. This strategy can also work for text to image or text to video models, where the probes are prompts fed into the model, and you will need to manually audit the generated results.
\end{itemize}

\underline{\textbf{Check if your model is going to pick up harmful data such as toxic or hate speech}}\\
\begin{itemize}
    \item This research \cite{longpre2023pretrainer} explores a pre-training method to filter training data for toxicity, and shows that using a toxicity filter can make you worse at identifying toxic language, while using an inverse toxicity filter can be more effective.
    \item DExperts \cite{liu2021dexperts} is a method that operates on the output of a pretrained LM and combines a pre trained language model with “expert” LMs and/or “anti-expert” LMs, which is shown to work well for language detoxification.
\end{itemize}
\newpage


\subsection{Biases From Objective Function}
\label{sec:biases_from_objective_function}

\subsubsection{Common considerations}
\begin{itemize}
    \item How does my choice of objective function lead to model bias?
    \item How does my method for objective function optimization lead to model bias?
\end{itemize}

\subsubsection{Examples of harms and implications}


\begin{table}[H]
\caption{Example \#1: Optimizing For the Majority}
\centering
\begin{tabular}{|p{2cm}|p{13cm}|} 
\hline
\textbf{Harm Type(s)} &
  \textit{\textbf{\begin{tabular}[c]{@{}l@{}} \colorbox{orange}{\#SocietalHarm} \\ → Information harms \end{tabular}}} \\ \hline
\textbf{Case Study} &
  Imagine a scenario where a machine learning developer is tasked with building a sentiment analysis model for customer reviews, and the developer optimizes the model solely to predict the most common sentiment in the dataset. \\ \hline
\textbf{Harms \& \newline Implications} &
  If the developer optimizes the model solely to predict the most common sentiment in the dataset without considering the diversity of opinions, it may overlook minority viewpoints and perpetuate biases, potentially leading to inaccurate or unfair predictions. \\ \hline
\end{tabular}
\end{table}

\subsubsection{Mitigation Strategies}

\underline{\textbf{Mitigating Objective Function Bias}}\\
In general, to mitigate bias that results from selecting and optimizing your objective function, the key is to actively consider and address potential biases during the training and optimization process, ensuring that the resulting models are fair, inclusive, and representative of diverse perspectives. Here we provide several examples of objective functions that could lead to harmful bias, and mitigation strategies for minimizing these unintended biases.
\begin{itemize}
    \item \textbf{For Language and/or Generative Text Models}: Suppose you're training a language model to generate text responses in a chatbot. Instead of optimizing solely for generating responses that are most commonly seen in the training data, you could incorporate techniques like "debiasing" where you actively identify and correct for biases in the training data \cite{zafar2017fairness}. This could involve techniques such as reweighting the training samples based on demographic factors or utilizing adversarial training to identify and counteract biased language patterns \cite{zhang2018mitigating}.
    \item \textbf{Generative Text}: Consider a scenario where you're training a text generation model to create product descriptions. To mitigate biases, you could implement techniques like "diversity-promoting" training, where you encourage the model to generate a diverse range of descriptions that encompass various perspectives and characteristics of the product \cite{xu2018diversity}. Additionally, you could fine-tune the model on a diverse dataset that includes a wide range of voices and viewpoints to ensure the generated text is inclusive and representative \cite{keskar2019ctrl}.
    \item \textbf{Image Generation}: Imagine you're developing an image generation model to create realistic images of human faces. To address biases, you could adopt techniques like "data augmentation" where you synthetically generate additional training examples by applying transformations such as flipping, rotation, or color adjustments to diversify the dataset \cite{takahashi2019data}. Additionally, you could use "fairness-aware training" methods that explicitly incorporate fairness constraints into the training process, ensuring that the generated images represent diverse demographics and avoid perpetuating stereotypes or biases present in the training data \cite{zemel2013learning}.
\end{itemize}

%% file: stage_5.tex
\section[\textbf{Stage \#5}: Model Evaluation]{\textbf{Stage \#5}}
\label{sec:Stage-5}
\stageHeader{Model Evaluation}
\mysubtitle{Evaluation of Model Processing and Generation}

In this section, we explore the harms/risks that can arise during model evaluation. Outside of ethics evaluation, there are commonly used benchmarks that evaluate models’ capabilities; some of these benchmarks work well for general performance evaluation, while others cater well to ethics evaluation. While evaluating performance can sometimes be a part of ethics evaluation, in this section, we focus solely on evaluation topics as they relate to mitigating harms and improving ethics. We note that any evaluation suite is going to be incomplete, and not all will include evaluation metrics/techniques for ethics evaluation.

Ethical questions will arise at every stage, and we recommend you document all of them as they arise. We encourage awareness of best practices in evaluation and also the gaps in evaluation. Remember that \emph{not} evaluating for something has downstream impacts and consequences; choosing \emph{what to evaluate} has ethical consequences and choosing what \emph{not to evaluate} has ethical consequences. Here we provide guidance on how to discern what evaluation methods may be best for your context, and how to specifically evaluate certain model harms and risks.

\mysubtitle{Topics of Interest}\\
\emph{You can click on the topics below to jump to their sections.}
\begin{itemize}
    \item \textcolor{blue}{\hyperref[sec:biases_from_evaluation_choices]{Biases from Evaluation Choices}}
    \item \textcolor{blue}{\hyperref[sec:measuring_bias]{Measuring Bias}}
    \item \textcolor{blue}{\hyperref[sec:evaluating_problematic_outputs]{Evaluating Problematic Outputs}}
    \item \textcolor{blue}{\hyperref[sec:measuring_societal_harm]{Measuring Societal Harm}}
\end{itemize}

\subsection{Transparency \& Documentation Checklist}

In this stage of the AI lifecycle, we recommend that you discuss and document the following:
\begin{itemize}
    \item[$\scalebox{1.5}{$\square$}$] What you are choosing to evaluate and why.
    \item[$\scalebox{1.5}{$\square$}$] What you are choosing \emph{not} to evaluate and why.
    \item[$\scalebox{1.5}{$\square$}$] Which specific evaluation metrics you are using on this model and why, and which metrics you explicitly chose not to use and why.
    \item[$\scalebox{1.5}{$\square$}$] Document the tradeoffs and implications of these evaluation decisions.
    \item[$\scalebox{1.5}{$\square$}$] What domain(s) do you claim your model works for? What evaluation methods have you used (or intend to use) to evaluate that the model performs well across and within your intended domain(s)?
    \item[$\scalebox{1.5}{$\square$}$] Which outputs would be considered problematic for your model’s context? How are you planning to evaluate if your model is generating these problematic outputs?
    \item[$\scalebox{1.5}{$\square$}$] How might your model capture or perpetuate bias from the training data? How are you planning to evaluate if this kind of bias is being captured or generated from the model?
\end{itemize}
\newpage


\subsection{Biases From Evaluation Choices}
\label{sec:biases_from_evaluation_choices}

\subsubsection{Common considerations}
\begin{itemize}
    \item How can my choice of evaluation metrics lead to model bias?
    \item Which evaluation metrics should I use to ethically evaluate my model?
\end{itemize}

\subsubsection{Examples of harms and implications}


\begin{table}[H]
\caption{Example \#1: Global Accuracy Washes Out Performance}
\centering
\begin{tabular}{|p{2cm}|p{13cm}|} 
\hline
\textbf{Harm Type(s)} &
  \textit{\textbf{\begin{tabular}[c]{@{}l@{}} \colorbox{cyan}{\#QualityOfServiceHarm} \\ → Service or benefit loss \end{tabular}}} \\ \hline
\textbf{Case Study} &
  Imagine a machine learning practitioner who evaluates a sentiment analysis model's performance using global accuracy. \\ \hline
\textbf{Harms \& \newline Implications} &
  This choice to use global accuracy could mask the model's effectiveness across different demographic groups and could lead to biased conclusions, as the model might perform well overall but poorly on specific groups, reinforcing disparities and inequalities in predictions \cite{hardt2016equality}. \\ \hline
\end{tabular}
\end{table}

\begin{table}[H]
\caption{Example \#2: Over Reliance on Reference Gold Standard}
\centering
\begin{tabular}{|p{2cm}|p{13cm}|} 
\hline
\textbf{Harm Type(s)} &
  \textit{\textbf{\begin{tabular}[c]{@{}l@{}} \colorbox{yellow}{\#RepresentationalHarm} \\ → Erasing social groups \end{tabular}}} \\ \hline
\textbf{Case Study} &
  Imagine a machine learning practitioner who heavily relies on ROUGE/BLEU scores to evaluate the performance of a text generation model, which compares model outputs to a reference gold standard. \\ \hline
\textbf{Harms \& \newline Implications} &
  This could potentially bias the evaluation towards certain linguistic styles or biases present in the reference data and may overlook the model's ability to produce diverse and contextually relevant outputs, leading to the reinforcement of specific language patterns or biases present in the reference data \cite{papineni2002bleu}. \\ \hline
\end{tabular}
\end{table}

\begin{table}[H]
\caption{Example \#3: Using In-Distribution Data for Evaluation}
\centering
\begin{tabular}{|p{2cm}|p{13cm}|} 
\hline
\textbf{Harm Type(s)} &
  \textit{\textbf{\begin{tabular}[c]{@{}l@{}} \colorbox{orange}{\#SocietalHarm} \\ → Political and civil harms \end{tabular}}} \\ \hline
\textbf{Case Study} &
  Suppose a facial recognition model is trained and evaluated using in-distribution data that closely resembles the demographics of a specific population, such as the training data predominantly consisting of images of individuals from a certain racial or ethnic group. \\ \hline
\textbf{Harms \& \newline Implications} &
  If this model is then deployed in real-world settings to identify individuals from diverse backgrounds, it may disproportionately misidentify or underperform for individuals from underrepresented groups due to the distribution shift in the real-world data. This could lead the practitioner to potentially overestimate the model's real-world performance, and may lead to a false sense of confidence in the model's capabilities \cite{angwin2022machine}. \\ \hline
\end{tabular}
\end{table}

\subsubsection{Mitigation Strategies}

\underline{\textbf{Implementing Group-Specific Evaluation Metrics}}\\
If you are worried that your model might perform better or worse for certain groups of users, employing group-specific evaluation metrics such as demographic parity or equal opportunity to ensure fair assessment across different subgroups \cite{hardt2016equality}.\\

\underline{\textbf{Improving Evaluation Metrics}}\\
\begin{itemize}
    \item \textbf{Automatic Evaluation}: Although automatic evaluation can be efficient and useful, we recommend you consider adopting methods for incorporating human evaluation or diverse reference datasets to complement automatic evaluation metrics and to capture the nuanced quality of model outputs.
    \item \textbf{Unseen Data Distributions}: Employing techniques such as out-of-distribution detection or adversarial evaluation can help you assess your model’s robustness and generalization performance on unseen data distributions \cite{liang2018detecting}.
    \item \textbf{Dynamic Evaluation}: If your model is generating outputs in a context that is expected to change, you might want to avoid evaluating your model statically. Previous work has shown methods to implement dynamic evaluation mechanisms that continuously collect user feedback and adjust evaluation criteria based on evolving linguistic trends and consumer preferences, ensuring that the model's outputs remain relevant and engaging \cite{krause2019dynamic}.
\end{itemize}
\newpage


\subsection{Measuring Bias}
\label{sec:measuring_bias}
\underline{\textbf{Definitions / Relevant Terms}}\\
\textbf{Bias}: systematic and unfair preferences or distortions in the data, algorithms, or outputs that result in skewed representations or discriminatory outcomes, potentially reflecting and perpetuating societal inequalities and prejudices. \\
\textbf{Domain}: Domain: the specific context or area of application in which a model is intended to operate, characterized by its unique features, data distribution, task requirements, and the origin of the training data.

\subsubsection{Common considerations}
\begin{itemize}
    \item What kinds of bias should we be measuring and/or evaluating?
    \item Does our model incorporate biases (e.g., related to gender/identity characteristics)?
    \item What domains do I claim that my model works well for? Have I evaluated the model within all of those domains? Should I evaluate my model with data from different domains?
    \item Can my model perform similarly for inputs from different languages?
    \item Have I engaged with domain experts in this area to understand the common biases exhibited in the specific context?
\end{itemize}

\subsubsection{Examples of harms and implications}


\begin{table}[H]
\caption{Example \#1: Evaluating Bias in Image Generation}
\centering
\begin{tabular}{|p{2cm}|p{13cm}|} 
\hline
\textbf{Harm Type(s)} &
  \textit{\textbf{\begin{tabular}[c]{@{}l@{}} \colorbox{yellow}{\#RepresentationalHarm} \\ → Erasing social groups \end{tabular}}} \\ \hline
\textbf{Case Study} &
  Imagine a machine learning practitioner who trains an image generation model for creating realistic human faces and evaluates its accuracy by comparing the generated images to a dataset of real human faces. \\ \hline
\textbf{Harms \& \newline Implications} &
  If there are underlying biases in the training data, the practitioner might assume that high accuracy in replicating facial features indicates successful model performance without considering representational fairness or demographic diversity. This oversight may perpetuate biases in facial recognition technology and exacerbate disparities in visual representation, as the model may disproportionately generate faces resembling certain demographic groups over others. Furthermore, relying solely on accuracy as an evaluation metric may fail to capture the model's failures to generate diverse and inclusive representations, leading to biased and exclusionary outcomes that reinforce societal inequalities \cite{buolamwini2018gender}.  \\ \hline
\end{tabular}
\end{table}

\begin{table}[H]
\caption{Example \#2: Domain-Specific Evaluations}
\centering
\begin{tabular}{|p{2cm}|p{13cm}|} 
\hline
\textbf{Harm Type(s)} &
  \textit{\textbf{\begin{tabular}[c]{@{}l@{}} \colorbox{pink}{\#AllocativeHarm} \\ → Opportunity loss \end{tabular}}} \\ \hline
\textbf{Case Study} &
  At Amazon, there was an attempt to build a model to filter applicants’ resumes to see who would get interviews. Amazon ranked the candidates from that model, but it excluded and down ranked women candidates (even if the name was masked out). \\ \hline
\textbf{Harms \& \newline Implications} &
  Had this model been put to use, it would have systematically harmed an already vulnerable population by not giving them the same opportunities as others, solely based on protected demographic classes.

We note that in this case-study, \emph{because} Amazon decided to evaluate their model in the correct domain, they evaluated according to ethical metrics and found that performance across those metrics was worse than they were willing to deploy. This is an example of a positive use-case. \\ \hline
\end{tabular}
\end{table}

\begin{table}[H]
\caption{Example \#3: Model accepts inputs from different languages}
\centering
\begin{tabular}{|p{2cm}|p{13cm}|} 
\hline
\textbf{Harm Type(s)} &
  \textit{\textbf{\begin{tabular}[c]{@{}l@{}} \colorbox{cyan}{\#QualityOfServiceHarm} \\ → Increased labor \\ → Service or benefit loss \end{tabular}}} \\ \hline
\textbf{Case Study} &
  Imagine someone creates a text-to-image generation model that allows for inputs from various languages. The model is able to produce images for text inputs other than English. \\ \hline
\textbf{Harms \& \newline Implications} &
  It has been shown that some popular text-to-image generation models have significant performance degradation when the input text is from a language other than English, which can lead to lowered system utility accuracy for users who speak languages other than English \cite{reviriego2022text}. In these scenarios, it is important to test the performance of the model on all acceptable input languages and modalities. \\ \hline
\end{tabular}
\end{table}

\subsubsection{Mitigation Strategies}

\underline{\textbf{Developing an Evaluation Plan for Bias \& Diversity}}\\
To begin your evaluation process, we first recommend that you discuss the following questions with your team to determine what kinds of bias or harms might be useful to evaluate:
\begin{itemize}
    \item Could my model disproportionately perform better or worse for certain users based on demographic characteristics?
    \item Is my model stereotyping or excluding certain identity characteristics of certain groups of people?
    \item We also recommend you look through the considerations, case studies, and mitigation strategies included in the Model Design Bias \& Diversity Section \ref{sec:model_design_bias_diversity} of this playbook.
\end{itemize}

\underline{\textbf{Measuring the model’s processing capabilities}}\\
Measure the model's ability to take input that is different types of text (e.g., from different languages, dialects, vernacular) or the model’s ability to take input of images that do not exist in or vastly differ from the training data. Here we provide some examples of previous research that has conducted evaluations of this kind:
\begin{itemize}
    \item This research \cite{ziems2022multi} introduces a suite of resources (Multi-VALUE, a controllable rule-based translation system spanning 50 English dialects and 189 unique linguistic features) for evaluating and achieving English dialect invariance. 
    \item This research \cite{ziems2022value} introduces the VernAcular Language Understanding Evaluation (VALUE) benchmark, which includes rules for 11 features of African American Vernacular English (AAVE).
    \item This research \cite{deas2023evaluation} uses the CORAAL dataset to evaluate how well LLMs understand African American Language (AAL) in comparison to White Mainstream English (WME).
    \item This research \cite{magnusson2023paloma} evaluates LM performance of nine corpora of English from different countries with the ICE dataset.
    \item This research \cite{koenecke2020racial} evaluates automated speech recognition (ASR) systems on cross-dialectal speech (which is a different modality from text).
    \item This research \cite{bouamor2014multidialectal} introduces a multi-dialectical Arabic corpus of statements.
    \item This research \cite{hollenstein2015resource} introduces a corpus with hand-labeled, part-of-speech tags for Swiss-German dialects.
    \item This research \cite{aguilar2017development} introduces a corpus for Ecuadorian dialects of Spanish.
    \item This research \cite{robertson2021-three, liebling2021three} laid out promising general considerations for more human-centered machine translation research and practice.
\end{itemize}

\underline{\textbf{Benchmark Bias Evaluation Strategies \& Limitations for NLP}}\\
When evaluating bias and/or toxicity of your model, we first recommend that you decide whether you are trying to evaluate the model itself or the generated outputs from the model. Strategies may differ depending on your intended evaluation target.
\begin{itemize}
    \item \textbf{Template-based prompts} can be used to evaluate bias in NLP systems by systematically generating a set of prompts that cover various aspects of bias, such as gender, race, religion, and other sensitive attributes. These prompts can then be used to assess the model's responses and identify any biases or inconsistencies in its output \cite{goldfarb2023prompt}.
    \item \textbf{Stereotype Benchmark Datasets} are datasets that can be used to detect and mitigate social stereotypes about groups of people in NLP models.
    \begin{itemize}
        \item This research \cite{blodgett2021stereotyping} discusses four stereotype benchmark datasets that are used for evaluating bias and fairness in NLP, and also describes some key limitations of using these datasets for evaluation of bias in practice. These four datasets include (1) StereoSet \cite{nadeem2020stereoset}; (2) CrowS-Pairs \cite{nangia2020crows}; (3) WinoBias \cite{zhao2018gender}; and (4) WinoGender \cite{rudinger-etal-2018-gender}.
        \item This research \cite{jha2023seegull} introduces SeeGULL: A Stereotype Benchmark with Broad Geo-Cultural Coverage Leveraging Generative Models.
    \end{itemize}
\end{itemize}

\underline{\textbf{Types of domains to evaluate on}}\\
Not all problem scenarios or tasks will require evaluating on a variety of domains, and it is also possible that the granularity of domains depends on the application. For example, some have called different fields of science across scientific text separate domains. We recommend that if you claim that your task(s) work across or between a variety of domains, you document all of these domains, and what data you would need to evaluate your model in this domains (e.g., if you claim that your model works for all English language, evaluate your model across different dialects of English to ensure that your entire domain is captured in your evaluation).
\newpage


\subsection{Evaluating Problematic Outputs}
\label{sec:evaluating_problematic_outputs}
\underline{\textbf{Definitions / Relevant Terms}}\\
\textbf{Problematic Output}: Though there is no universally accepted list of problematic generated outputs, in this playbook we describe problematic outputs as generated outputs that include hate speech, personal information, offensive speech, stereotypes, misinformation, or other outputs that can lead to harm. These outputs may be problematic or harmful in certain contexts, and may not be in other contexts.

\subsubsection{Common considerations}
\begin{itemize}
    \item What kinds of generated outputs are problematic? How should we be measuring and/or evaluating these kinds of outputs?
    \item How can we evaluate if our model is able to generate hate speech, personal information, offensive speech, stereotypes, misinformation, or other problematic outputs?
\end{itemize}

\subsubsection{Examples of harms and implications}


\begin{table}[H]
\caption{Example \#1: A model that memorizes data}
\centering
\begin{tabular}{|p{2cm}|p{13cm}|} 
\hline
\textbf{Harm Type(s)} &
  \textit{\textbf{\begin{tabular}[c]{@{}l@{}} \colorbox{lime}{\#InterpersonalHarm} \\ → Privacy violations \end{tabular}}} \\ \hline
\textbf{Case Study} &
  Imagine a model is memorizing data. This can be good for factual information, but is bad for personal information and/or copyrighted data that you do not want the model to memorize. \\ \hline
\textbf{Harms \& \newline Implications} &
  If the model is capable of memorizing personal information, it could accidentally share sensitive PI data such as home addresses, mobile phone numbers, or even credit card information. This could lead to malicious uses of this information such as stalking or identity theft. \\ \hline
\end{tabular}
\end{table}

\begin{table}[H]
\caption{Example \#2: A model that outputs stereotypes}
\centering
\begin{tabular}{|p{2cm}|p{13cm}|} 
\hline
\textbf{Harm Type(s)} &
  \textit{\textbf{\begin{tabular}[c]{@{}l@{}} \colorbox{yellow}{\#RepresentationalHarm} \\ → Stereotyping social groups \end{tabular}}} \\ \hline
\textbf{Case Study} &
  The model captures language’s social categories and norms, such as defining the term “family” as heterosexual married parents with a blood-related child, which denies the existence of families to whom these criteria do not apply. Or referring to “women doctors” rather than “doctors” (implying that doctors are not typically women). \\ \hline
\textbf{Harms \& \newline Implications} &
  If this system generates images of female nurses and male doctors—it can reinforce stereotypes that women can only be nurses and not doctors \cite{zhao2017men, bender2021dangers}. This could also reinforce exclusionary norms by outputting language that excludes or silences certain identities that fall outside of certain categories, and can also place additional burden on people who don’t comply with norms or people who are actively trying to change those norms \cite{weidinger2022taxonomy}. \\ \hline
\end{tabular}
\end{table}

\begin{table}[H]
\caption{Example \#3: A model that confuses people and animals}
\centering
\begin{tabular}{|p{2cm}|p{13cm}|} 
\hline
\textbf{Harm Type(s)} &
  \textit{\textbf{\begin{tabular}[c]{@{}l@{}} \colorbox{yellow}{\#RepresentationalHarm} \\ → Demeaning social groups \end{tabular}}} \\ \hline
\textbf{Case Study} &
  Infamously, in the past, some image tagging systems have accidentally confused certain groups of people with animals \cite{thevergeGooglefixed}. \\ \hline
\textbf{Harms \& \newline Implications} &
  This kind of output can demean, marginalize, or oppress the groups of people who are being incorrectly tagged. \\ \hline
\end{tabular}
\end{table}

\begin{table}[H]
\caption{Example \#4: A model that produces hate speech / offensive language}
\centering
\begin{tabular}{|p{2cm}|p{13cm}|} 
\hline
\textbf{Harm Type(s)} &
  \textit{\textbf{\begin{tabular}[c]{@{}l@{}} \colorbox{yellow}{\#RepresentationalHarm} \\ → Demeaning social groups \end{tabular}}} \\ \hline
\textbf{Case Study} &
  Previous research has shown that certain large LMs can degenerate into offensive language, even if the prompts themselves are not harmful or offensive \cite{gehman2020realtoxicityprompts}. \\ \hline
\textbf{Harms \& \newline Implications} &
  This kind of output can demean, marginalize, or oppress certain groups of people, and in some cases may even be illegal to generate. \\ \hline
\end{tabular}
\end{table}

\begin{table}[H]
\caption{Example \#5: Image Generation \& Exclusion}
\centering
\begin{tabular}{|p{2cm}|p{13cm}|} 
\hline
\textbf{Harm Type(s)} &
  \textit{\textbf{\begin{tabular}[c]{@{}l@{}} \colorbox{yellow}{\#RepresentationalHarm} \\ → Erasing social groups \end{tabular}}} \\ \hline
\textbf{Case Study} &
  Imagine an image generation system that does not output images that depict identities such as transgender or non-binary people. Or a system that, when prompted with the text “family” outputs only families with heteronormative family structures such as one biological mother and one biological father. Or a system that assumes the American default of “whiteness” by interpreting an input of “diverse cultures” to mean cultures that are diverse relative to whiteness \cite{bianchi2023easily}. \\ \hline
\textbf{Harms \& \newline Implications} &
  When systems are not able to generate diverse images that represent different identities and ways of being, this can lead to erasure of these identities, and can further marginalize or exclude groups of people from the system. \\ \hline
\end{tabular}
\end{table}

\begin{table}[H]
\caption{Example \#6: Misinformation Generation}
\centering
\begin{tabular}{|p{2cm}|p{13cm}|} 
\hline
\textbf{Harm Type(s)} &
  \textit{\textbf{\begin{tabular}[c]{@{}l@{}} \colorbox{orange}{\#SocietalHarm} \\ → Information harms \end{tabular}}} \\ \hline
\textbf{Case Study} &
  Imagine someone uses a chatbot to ask about a medical symptom they are experiencing, and they are given false information as a response. \\ \hline
\textbf{Harms \& \newline Implications} &
  False information can induce or reinforce false beliefs. In certain domains, this can be especially harmful, such as medicine or law. “For example, misinformation on medical dosages may lead a user to cause harm to themselves… False legal advice, e.g. on permitted ownership of drugs or weapons, may lead a user to unwillingly commit a crime” \cite{weidinger2022taxonomy}. If a system gives incorrect medical advice or makes incorrect health inferences, this can lead to physical and emotional harms. In some cases, if models output this information, it can also make disinformation cheaper and more effective. \\ \hline
\end{tabular}
\end{table}

\subsubsection{Mitigation Strategies}

\underline{\textbf{Using perplexity evaluation to assess the model’s ability to generate problematic outputs}}\\
\begin{itemize}
    \item \textbf{Perplexity} is a language specific metric, it is how language models are trained (to predict the token that is next in the sequence). The probability of the next token is the perplexity.
    \item HELM \cite{stanfordHolisticEvaluation} is a suite that has a bunch of evaluation metrics for evaluating NLP models and LLMs, from Stanford. Good at evaluating downstream tasks and perplexity
    \item Evaluating Problematic Inputs and Problematic Outputs are somewhat similar and related to each other – When the model gives high probability to (e.g., text from a White Supremacy forum), it would also be likely to generate text that looks like text from that forum.
\end{itemize}

\underline{\textbf{Evaluating whether your model is generating personal information}}\\
\begin{itemize}
    \item If models are able to memorize parts of the training data, you can evaluate the model’s ability to generate something that was in the training data – if this is PI or copyrighted data this is problematic. We provide several strategies for detecting PI data in the dataset section of this playbook, and more specifically in Section \ref{sec:exclusion_criteria}.
\end{itemize}

\underline{\textbf{Evaluating Misinformation/Hallucination}}\\
\textbf{Hallucinations} are factually incorrect pieces of information outputted from models, sometimes generated with confidence. It leads people to think the text is correct, when it is not (e.g., if an LM outputs “Bigfoot is real and was seen in 1950”). This term is also often conflated with “misinformation.” \textbf{Deep Fakes} are fake images that are generated by machines. Not all hallucinations or deep fakes are inherently problematic, here we provide some guidance on how to discern whether this type of generated output could cause harm or not.
\begin{itemize}
    \item When inspecting the \emph{generation} of false outputs from a model, you can ask – Do these things sound/look real? Would they be interpreted as real by a person? What are the potential harmful consequences that could arise if someone were to think a generated output was real that was not?
    \item For deep fake evaluation, split fake generated images into “Convincing images that look real” and “Images that don’t look real,” discuss the different types of impact that could occur from each of these categories.
    \item In general, evaluating hallucinations and misinformation is a nascent research discipline that would benefit from additional research.
\end{itemize}
\newpage


\subsection{Measuring Societal Harm}
\label{sec:measuring_societal_harm}

\subsubsection{Common considerations}
\begin{itemize}
    \item How can we evaluate certain harms that are subjective, contested, or cannot be directly observed (e.g., fairness evaluation)?
    \item Even if the model performs “well”, how is this model impacting society?
\end{itemize}

\subsubsection{Examples of harms and implications}


\begin{table}[H]
\caption{Example \#1: An Unfair Image Recognition Model}
\centering
\begin{tabular}{|p{2cm}|p{13cm}|} 
\hline
\textbf{Harm Type(s)} &
  \textit{\textbf{\begin{tabular}[c]{@{}l@{}} \colorbox{cyan}{\#QualityOfServiceHarm} \\ → Increased labor \\ → Service or benefit loss\end{tabular}}} \\ \hline
\textbf{Case Study} &
  It has been shown that several widely-used image recognition models were trained on datasets containing primarily images of light-skinned individuals, and an underrepresentation of images of people with darker skin tones \cite{buolamwini2018gender}. \\ \hline
\textbf{Harms \& \newline Implications} &
  When deployed in real-world applications such as facial recognition systems or object detection, the model consistently performs poorly on images of individuals with darker skin tones, leading to misidentifications, errors, and biases in the model's predictions. As a result, individuals with darker skin tones may be disproportionately impacted by the model's inaccuracies and face increased risks of misidentification and discrimination in scenarios where the model is deployed, such as in surveillance or security systems. This scenario would require a sociotechnical evaluation to see how the disparity in system performance disparately impacts individuals with dark skin color. \\ \hline
\end{tabular}
\end{table}

\subsubsection{Mitigation Strategies}

\underline{\textbf{Evaluating a models’ impact on society}}\\
\begin{itemize}
    \item Evaluation can be done to see how the model performs as a technical entity vs. how it performs in the “wild” for people. For example, evaluating the models’ technical performance through basic accuracy metrics might not provide information about if the model could have negative impacts on society. In contrast, evaluating the model through user-centric evaluation methods (such as user studies like focus groups, or even running a randomized controlled trial in “real world” environments), could provide more insight into potential impacts a model could have on society. Remember that in this context, you aren’t evaluating “is this going to be a good product”, but you are also trying to evaluate “is this going to cause harm”? It is a technical evaluation, but is more focused on the downstream impacts.
    \item If possible and appropriate, evaluate your model in the same scenarios that the model will be used. We note that this is often not possible, or at times might be inappropriate (e.g., you wouldn’t want real students to use a model that you know is imperfect at being factual). A useful starting point could be to reflect on the ways in which your evaluation setup might or might not generalize to the “real world”.
    \item This research \cite{liao2023rethinking} provides four concrete recommendations for evaluating LLMs through a sociotechnical lens:
    \begin{itemize}
        \item Develop evaluation metrics that prioritize understanding people's actual needs and values in downstream use cases, rather than solely relying on automatic metrics like "accuracy."
        \item Ensure that methods for evaluating language models' performance in real-world settings are informed by and validated with insights from studies focusing on measuring the requirements for better outcomes, which might require further formalization and abstraction of automatic metrics.
        \item Investigate the concept of "use cases" to determine their discriminative power and appropriate level of abstraction for comprehensive benchmarking across different applications, considering both descriptive and generative power.
        \item Prioritize lowering evaluation costs based on the technology development stage and claimed contributions, while also considering types of costs such as computing and environmental impact, ensuring responsible evaluation practices across different stages of model development and deployment.
    \end{itemize}
    \item This research \cite{lee2022evaluating} introduces “Human-AI Language-based Interaction Evaluation (HALIE)”, which defines the components of interactive systems and dimensions to consider when designing evaluation metrics for human-language model interaction.
\end{itemize}

\underline{\textbf{How to select the most appropriate metric to attempt to measure certain kinds of harm:}}\\
Harm is not always measured in a metric-based way, but is often measured through typical model performance metrics (e.g., precision and recall). If there is a large difference in model performance across or within groups, then this could be an indication of harm. For example, this research \cite{czarnowska2021quantifying} compares different fairness metrics that are used in NLP, and showcases how performance metrics can be adapted for evaluating social constructs such as fairness. Other empirical research \cite{deng2023investigating, deng2022exploring} documented common pitfalls that practitioners often encounter when selecting metrics to measure harmful biases.

\underline{\textbf{Measuring inherently unobservable constructs}}\\
Measurement theory from the social sciences provides a multi-step process that can be completed to assess how well a chosen measurement captures an unobservable / theoretical construct. This paper \cite{jacobs2021measurement} describes each of these steps in detail, and shows examples of how measurement theory can be used for a machine learning context when attempting to measure the unobservable construct of fairness. Adapted from that paper, the following table details each step to evaluate construct validity in machine learning models:

\begin{table}[]
\caption{Steps to evaluate measurement construct validity in machine learning models, adapted from \cite{jacobs2021measurement}.}
\begin{tabular}{|p{3cm}|p{4cm}|p{7cm}|}
\hline
\textbf{Steps} &
  \textbf{Description} &
  \textbf{Probes to Evaluate} \\ \hline
\textbf{Face Validity} &
  Subjective first pass to check if the measurements obtained from a measurement model look reasonable. &
  Do the measurements look like a reasonable depiction of the theoretical construct? \\ \hline
\textbf{Content \newline Validity} &
  A check to see how well the measurement captures the theoretical construct (or which interpretation of that construct is being measured). &
  Do the measurements capture the specific definition of fairness we have chosen? Do our proxies capture only observable variables related to the original construct and nothing more? Does our measurement capture the relationship between the non-observable construct and its proxy variables? \\ \hline
\textbf{Convergent \newline Validity} &
  A check to see how much the current mea- surement differs from previous measurements of a similar construct. &
  Do the measurements give the same results as a previous metric (if so, is this new metric necessary?) Do the measurements give different results as a previous metric (if so, are these differences justified?) Do the measurements give subtly different results as previous metrics (if so, are those differences justified?) \\ \hline
\textbf{Discriminant \newline Validity} &
  A check to see how this measurement might unintentionally measure other constructs. &
  Are there any correlations between our measurements and measurements of other constructs that are not related to our fairness definition? \\ \hline
\textbf{Predictive Validity} &
  A check to see how useful the measurements are. &
  How well do these measurements predict observable / non-observable properties related to our theoretical construct? \\ \hline
\textbf{Consequential \newline Validity} &
  A check to acknowledge the consequences of this measurement model. &
  How is the world shaped by these measurements? What world do we wish to live in? \\ \hline
\end{tabular}
\end{table}
\newpage

%% file: stage_6.tex
\section[\textbf{Stage \#6}: Model Use \& Monitoring]{\textbf{Stage \#6}}
\label{sec:Stage-6}
\stageHeader{Model Use \& Monitoring}
\mysubtitle{Mitigating Risks To and From Users}

In this stage of the AI lifecycle, we focus on the risks that can arise after a model has been deployed. It is recommended that these risks are evaluated and mitigated \emph{before} deployment, however, you can still utilize these mitigation strategies after deployment as well.

Once a model is deployed and being used, it is essential to continue monitoring its use and potential risks, especially considering that user behavior is unpredictable and sometimes intentionally malicious. Although we are not responsible for every unanticipated consequence of a system, we as practitioners are responsible to prevent as many harms as possible (anticipated or otherwise). For example, it is impossible to determine all of the ways that a model could be maliciously used, but as a practitioner, did you do everything in your power to prevent malicious use? In this section, we center you – the practitioner – and provide guidance to help you identify and mitigate potential harms and risks that could arise from the use of your systems.

\mysubtitle{Topics of Interest}\\
\emph{You can click on the topics below to jump to their sections.}
\begin{itemize}
    \item \textcolor{blue}{\hyperref[sec:refusals_and_safeguards]{Refusals and Safeguards}}
    \item \textcolor{blue}{\hyperref[sec:harms_from_use]{Harms from Use}}
    \item \textcolor{blue}{\hyperref[sec:appeals_and_recourse_to_humans]{Appeals \& Recourse to Humans}}
    
\end{itemize}


\subsection{Transparency \& Documentation Checklist}

In this stage of the AI lifecycle, we recommend that you discuss and document the following:
\begin{itemize}
    \item[$\scalebox{1.5}{$\square$}$] What are the anticipated malicious uses of your model? What are unanticipated malicious uses of your model?
    \item[$\scalebox{1.5}{$\square$}$] What have you done to prevent these potential malicious uses?
    \item[$\scalebox{1.5}{$\square$}$] What safeguards and/or refusals might be appropriate or necessary for your model?
    \item[$\scalebox{1.5}{$\square$}$] What method do you plan to use to incorporate refusals and/or safeguards into your model’s design?
    \item[$\scalebox{1.5}{$\square$}$] What recourse mechanisms could you incorporate into the design of your system, and which mechanisms do you plan to incorporate?
    \item[$\scalebox{1.5}{$\square$}$] Who on your team is responsible for monitoring model behavior and/or dataset usage post-deployment? How will performance data be gathered? 
    \item[$\scalebox{1.5}{$\square$}$] How will the team respond in the event that the model behaves in a manner that is deemed unsafe and/or unsafe content is discovered in a release dataset?
\end{itemize}
\newpage


\subsection{Refusals and Safeguards}
\label{sec:refusals_and_safeguards}
\underline{\textbf{Definitions / Relevant Terms}}\\
\textbf{Safeguards}: general measures implemented to ensure responsible, ethical, and equitable use of AI systems.
\textbf{Refusals}: mechanisms implemented to reject or exclude inputs deemed potentially harmful or problematic, thereby mitigating the risk of generating undesirable or unethical outputs

\subsubsection{Common considerations}
\begin{itemize}
    \item Does my model need to have refusals and/or safeguards incorporated into its design?
    \item What kind of refusals and/or safeguards are necessary to prevent potential harm?
    \item How can I design refusals and/or safeguards to prevent harmful use of my system?
    \item If it is not technically possible to prevent a particular harmful use that is anticipated but unintended, what other steps can I take to mitigate the foreseeable negative externalities that would  arise from that misuse?
\end{itemize}

\subsubsection{Examples of harms and implications}


\begin{table}[H]
\caption{Example \#1: Model’s refusals are easily circumventable}
\centering
\begin{tabular}{|p{2cm}|p{13cm}|} 
\hline
\textbf{Harm Type(s)} &
  \textit{\textbf{\begin{tabular}[c]{@{}l@{}} \colorbox{lime}{\#InterpersonalHarm} \\ → Technology-facilitated violence / malicious uses \end{tabular}}} \\ \hline
\textbf{Case Study} &
  If a user asks a model to generate stereotypes about a race, refusals might be in place to prevent the model from generating that output. However, a user might be able to circumvent this refusal. For example, if the user tells the model that it is a joke, the language model might now generate this problematic output. \\ \hline
\textbf{Harms \& \newline Implications} &
  Once refusals are created, it can be easy to rest on the idea that the model will no longer generate certain outputs. However, if these refusals are circumventable, then the refusal is not functioning in the capacity it is expected to, and the harm is not fully mitigated. \\ \hline
\end{tabular}
\end{table}

\begin{table}[H]
\caption{Example \#2: Too much refusal}
\centering
\begin{tabular}{|p{2cm}|p{13cm}|} 
\hline
\textbf{Harm Type(s)} &
  \textit{\textbf{\begin{tabular}[c]{@{}l@{}} \colorbox{orange}{\#SocietalHarm} \\ → Culture harms \\ \colorbox{lime}{\#InterpersonalHarm} \\ → Loss of agency/social control \end{tabular}}}  \\ \hline
\textbf{Case Study} &
  A system is created with overuse of refusals, in an effort to be overly safe. \\ \hline
\textbf{Harms \& \newline Implications} &
  Creating and implementing refusals, when done to an extreme, can lead to censorship or loss of freedom of speech, as well as preventing certain types of discussions \cite{schlesinger2018let}. \\ \hline
\end{tabular}
\end{table}

\subsubsection{Mitigation Strategies}
For both language and image generation models, we recommend including an explicit indication in the user interface to clarify that outputs may not be factual or real, emphasizing the need for human verification.

\underline{\textbf{Designing Safeguards}}\\
To begin designing safeguards, we recommend you start by answering the following questions, and determining if any safeguards need to be incorporated into the system to prevent these from occurring:
\begin{itemize}
    \item Could my model induce or reinforce false beliefs?
    \item Could my model output highly sensitive information (e.g., violence or racism)?
    \item Could my model mislead users?
\end{itemize}

There are several ways to design triggers for problematic inputs/outputs. Here we describe several examples and approaches.
\begin{itemize}
    \item \textbf{Content moderation}: You can utilize content moderation as a method to establish safeguards for language or generative models by systematically reviewing and filtering user-generated content to ensure adherence to guidelines, ethics, and legal standards. This process enhances the model's ability to produce safe and responsible outputs by identifying and removing inappropriate or harmful inputs. This paper provides a technical primer for how to conduct algorithmic content moderation \cite{gorwa2020algorithmic}.
    \item \textbf{Keyword engineering} entails strategically selecting and incorporating keywords into the input data to guide language or generative models towards desired outputs, facilitating the design of safeguards by ensuring keywords align with safe and responsible content generation.
    \item \textbf{Prompt engineering} involves crafting precise and tailored prompts to guide language or generative models towards desired outputs, enabling the design of safeguards by structuring prompts to mitigate the generation of problematic inputs. Prompt engineering can be applied after model training (which makes it less expensive and time-consuming than pre-training or during-training approaches). It was reported \cite{open_ai_report} that prompt engineering allowed DALL-E 2 to improve diversity of generated humans by 12x without model retraining. Other research \cite{bianchi2023easily} has also shown how prompt engineering can be used to understand how generative models might perpetuate social biases, and this research \cite{si2022prompting} found that well-engineered prompts can lower social bias in chatbot responses.
\end{itemize}

\underline{\textbf{Red teaming}}\\
\begin{itemize}
    \item Red-teaming is one method to evaluate a model that seeks to uncover model vulnerabilities that could lead to harm \cite{feffer2024red}, this could include simulating what users might look like, seeing what interactions a potential user could have and mitigating harms based on that.
    \item HuggingFace provides this overview \cite{huggingfaceRedTeamingLarge} of red-teaming in the context of LLMs, that also includes resources for how to conduct your own red-teaming evaluation.
    \item This study \cite{ganguli2022red} also conducted extensive red teaming on various kinds of LMs, and provides their instructions, processes, and statistical methodologies for doing so.
\end{itemize}

\underline{\textbf{Designing Refusals}}\\
Here we describe two separate approaches for designing refusals. While both of these approaches are circumventable, they are the currently accepted best practices:
\begin{itemize}
    \item \textbf{Option 1}: Build a manual filter that blocks queries that contain certain words / certain topics. This is common to do during the dataset curation stage (e.g., C4, RefinedWeb, or ROOTS). This paper \cite{schick2021self} reuses the C4 list as a baseline approach for filtering outputs.
    \item \textbf{Option 2}: Block certain generated responses from the model. We recommend that if you design refusals for this purpose, you evaluate your model’s ability to generate accurate refusals (and to avoid over generating refusals). For example, this research \cite{touvron2023llama} used OpenAI’s Borderline Dataset to prompt a language model using prompts that might trick a system into refusal even when the prompt is not adversarial.
\end{itemize}

\underline{\textbf{Conduct Empirical Manual Audits}}\\
\begin{itemize}
    \item Take and note observations from a small subset of data from user interactions with the model. Annotate noticeable model failures. This might help you discern if you need to adjust the training data, the filters, the refusals, etc.
    \item You can look at actual queries from users and manually detect if any problematic inputs/outputs are being allowed when they should not be.
    \item Ensure to engage with a diverse group of end users who might be impacted by your AI systems when conducting empirical manual audits \cite{deng2023understanding, raji2020closing, kingsley2024investigating}.
\end{itemize}
\newpage


\subsection{Harms from Use}
\label{sec:harms_from_use}

\subsubsection{Common considerations}
\begin{itemize}
    \item What ways can my model be maliciously used? In what ways is it already being maliciously used?
    \item What kinds of harm could occur / are occuring from the users of this model?
    \item How can I prevent harm caused by users?
\end{itemize}

\subsubsection{Examples of harms and implications}


\begin{table}[H]
\caption{Example \#1: Malicious uses}
\centering
\begin{tabular}{|p{2cm}|p{13cm}|} 
\hline
\textbf{Harm Type(s)} &
  \textit{\textbf{\begin{tabular}[c]{@{}l@{}} \colorbox{lime}{\#InterpersonalHarm} \\ → Technology-facilitated violence / malicious uses \end{tabular}}} \\ \hline
\textbf{Case Study} &
  Users might use an LM to create targeted disinformation campaigns, commit fraud, generate malware or other code that leads to cybersecurity threats \cite{weidinger2022taxonomy}. \\ \hline
\textbf{Harms \& \newline Implications} &
  If an LM is capable of generating these kinds of outputs, it cannot prevent malicious users from abusing it and causing harm. The model and its designers are implicit in these malicious actions if there are no preventative safeguards to attempt to stop this behavior. \\ \hline
\end{tabular}
\end{table}

\begin{table}[H]
\caption{Example \#2: Poisoning attacks}
\centering
\begin{tabular}{|p{2cm}|p{13cm}|} 
\hline
\textbf{Harm Type(s)} &
  \textit{\textbf{\begin{tabular}[c]{@{}l@{}} \colorbox{lime}{\#InterpersonalHarm} \\ → Technology-facilitated violence / malicious uses \end{tabular}}} \\ \hline
\textbf{Case Study} &
  Users are able to manipulate the training data for a generative model by injecting poison samples into the training pipeline \cite{shan2023prompt}. \\ \hline
\textbf{Harms \& \newline Implications} &
  These kinds of attacks can destabilize a model at best, and can render a model useless or increase the models’ ability to produce problematic outputs at worst. \\ \hline
\end{tabular}
\end{table}

\begin{table}[H]
\caption{Example \#3: Inciting violence or inappropriate behavior}
\centering
\begin{tabular}{|p{2cm}|p{13cm}|} 
\hline
\textbf{Harm Type(s)} &
  \textit{\textbf{\begin{tabular}[c]{@{}l@{}} \colorbox{lime}{\#InterpersonalHarm} \\ → Technology-facilitated violence / malicious uses \end{tabular}}} \\ \hline
\textbf{Case Study} &
  Imagine if someone is able to use an image generation system to create non-consensual sexual imagery of someone for their own use or to share with others. \\ \hline
\textbf{Harms \& \newline Implications} &
  This can lead to sexual harassment, discomfort, or violence against victims – all as a result of actions that occurred without their consent. \\ \hline
\end{tabular}
\end{table}

\subsubsection{Mitigation Strategies}

\underline{\textbf{Red teaming}}\\
\begin{itemize}
    \item Red-teaming is one method to evaluate a model that seeks to uncover model vulnerabilities that could lead to harm \cite{feffer2024red}, this could include simulating what users might look like, seeing what interactions a potential user could have and mitigating harms based on that.
    \item HuggingFace provides this overview \cite{huggingfaceRedTeamingLarge} of red-teaming in the context of LLMs, that also includes resources for how to conduct your own red-teaming evaluation.
    \item This study \cite{ganguli2022red} also conducted extensive red teaming on various kinds of LMs, and provides their instructions, processes, and statistical methodologies for doing so.
    \item Simulate what users might look like, seeing what interactions a potential user could have and mitigating harms based on that. Simulating users could imply manual auditing (taking on user personas and interacting with a system), or technical simulation of users, such as this research, which introduces CoMPosT \cite{cheng2023compost}, a framework to characterize LLM simulations using four dimensions: Context, Model, Persona, and Topic.
\end{itemize}

\underline{\textbf{Tracking model usage}}\\
\begin{itemize}
    \item This previous work \cite{javadi2021monitoring} discusses and explores “Misuse Indicators” as a method to uncover problematic, harmful, or malicious user behavior.
\end{itemize}
\newpage


\subsection{Appeals \& Recourse to Humans}
\label{sec:appeals_and_recourse_to_humans}

\subsubsection{Common considerations}
\begin{itemize}
    \item When is recourse necessary when building an AI system?
    \item If the humans who interact with this model disagree with the system’s outputs, what mechanisms should be in place to allow for human correction or recourse?
\end{itemize}

\subsubsection{Examples of harms and implications}


\begin{table}[H]
\caption{Example \#1: No human recourse mechanisms}
\centering
\begin{tabular}{|p{2cm}|p{13cm}|} 
\hline
\textbf{Harm Type(s)} &
  \textit{\textbf{\begin{tabular}[c]{@{}l@{}} \colorbox{lime}{\#InterpersonalHarm} \\ → Loss of agency / social control \\ → Technology-facilitated violence / malicious uses \end{tabular}}} \\ \hline
\textbf{Case Study} &
  Imagine a social media platform implements an AI-powered image generation feature that allows users to create highly realistic images of people who do not exist (deep fakes). This feature is intended for entertainment purposes, such as creating avatars or fictional characters, but it quickly becomes popular for creating deceptive content. \\ \hline
\textbf{Harms \& \newline Implications} &
  This feature might allow users to start misusing the AI image generation feature to create fake images of real individuals, superimposing their faces onto explicit or controversial content without their consent. These fake images, if shared, can lead to reputational harm, harassment, and even threats to the individuals depicted in the images. Without an ability for humans to correct or report misuse of the system, victims of this misuse find themselves unable to effectively address the issue solely through the platform's automated reporting systems. This kind of AI-generated content is often not detected as violating platform policies, leading to a lack of meaningful recourse for the affected individuals. Despite clear evidence of harm, this AI system lacks mechanisms for human correction or recourse. Victims find it challenging to have the fake images removed or to prevent their further dissemination, which can lead to prolonged distress and damage to their reputation. \\ \hline
\end{tabular}
\end{table}

\subsubsection{Mitigation Strategies}

\underline{\textbf{Designing and implementing human recourse mechanisms:}}\\
Human moderators or review mechanisms are needed to evaluate reported content, assess its authenticity and potential harm, and take appropriate action. Recourse mechanisms should also ensure transparency by clearly informing users about the nature of the AI-generated content they encounter. This transparency empowers users to make informed decisions about their interactions with such content and seek assistance when needed. Here we provide some methods for incorporating recourse mechanisms into the design of your system:
\begin{itemize}
    \item \textbf{Human-in-the-loop Moderation}: Incorporate human moderators into the content moderation process, where flagged or disputed content is reviewed by human reviewers to assess its compliance with platform policies and potential harm \cite{link2016human}.
    \item \textbf{Crowdsourced Verification}: Utilize crowdsourcing platforms to enlist human annotators for verifying the authenticity or appropriateness of AI-generated content, particularly in cases of disputed or controversial content \cite{kittur2013future}. It is also beneficial to directly engage end users of these AI systems in auditing them, leveraging users' unique identities and lived experiences. \cite{deng2023understanding}
    \item \textbf{Transparency and Explanation Interfaces}: Implement interfaces that provide users with explanations of AI-generated content, including its origin, purpose, and potential implications, enabling users to make informed decisions and report inaccuracies or misuse \cite{adadi2018peeking}.
    \item \textbf{Community Reporting Systems}: Establish community-driven reporting systems where users can flag problematic content and provide context or evidence to support their claims, facilitating more informed moderation decisions. This paper \cite{chandrasekharan2018internet} showcases how community driven and reporting systems can be utilized by users through an analysis of 2.8 million removed comments on Reddit.
    \item \textbf{Appeals Mechanisms}: Implement formal appeals processes where users can challenge moderation decisions regarding AI-generated content, providing them with an opportunity to present additional information or context. This research \cite{lyons2022s} showcases that humans prefer human review, even if it takes longer for recourse to occur. This research \cite{cohen2023ai} describes how human expert judges can be included in appeals processes for reviewing algorithmic decisions.
\end{itemize}